\documentclass[english,a4paper]{revtex4}
\usepackage[latin1]{inputenc}
\usepackage{float}
\usepackage{graphicx}

\makeatletter

\providecommand{\tabularnewline}{\\}

\usepackage{float}

\makeatletter

\usepackage{float}

\makeatletter

\usepackage{float}

\newcommand{\rd}{{\rm d}}
\newcommand{\m}{{\rm m}}



\makeatother

\makeatother

\usepackage{babel}
\makeatother
\begin{document}

\title{Conditions for the cosmological viability of $f(R)$ dark energy
models\\
 }

\author{Luca Amendola}

\affiliation{INAF/Osservatorio Astronomico di Roma, Via Frascati 33\\
 00040 Monte Porzio Catone (Roma), Italy}

\author{Radouane Gannouji}

\affiliation{LPTA, Universit\'{e} Montpellier II, CNRS, 34095 Montpellier Cedex
05, France}

\author{David Polarski}

\affiliation{LPTA, Universit\'{e} Montpellier II, CNRS, 34095 Montpellier Cedex
05, France}

\author{Shinji Tsujikawa}

\affiliation{Department of Physics, Gunma National College of Technology, Gunma
371-8530, Japan}

\date{\today{}}

\begin{abstract}
We derive the conditions under which dark energy models whose Lagrangian
densities $f$ are written in terms of the Ricci scalar $R$ are cosmologically
viable. We show that the cosmological behavior of $f(R)$ models can
be understood by a geometrical approach consisting in studying the
$m(r)$ curve on the $(r,m)$ plane, where $m\equiv Rf_{,RR}/f_{,R}$
and $r\equiv-Rf_{,R}/f$ with $f_{,R}\equiv{\rm d}f/{\rm d}R$. This
allows us to classify the $f(R)$ models into four general classes,
depending on the existence of a standard matter epoch and on the final
accelerated stage. The existence of a viable matter dominated epoch
prior to a late-time acceleration requires that the variable $m$
satisfies the conditions $m(r)\approx+0$ and ${\rm d}m/{\rm d}r>-1$
at $r\approx-1$. For the existence of a viable late-time acceleration
we require instead either (i) $m=-r-1$ , $(\sqrt{3}-1)/2<m\le1$
and ${\rm d}m/{\rm d}r<-1$ or (ii) $0\le m\le1$ at $r=-2$. These
conditions identify two regions in the $(r,m)$ space, one for the
matter era and the other for the acceleration. Only models with a
$m(r)$ curve that connects these regions and satisfy the requirements
above lead to an acceptable cosmology. The models of the type $f(R)=\alpha R^{-n}$
and $f=R+\alpha R^{-n}$ do not satisfy these conditions for any $n>0$
and $n<-1$ and are thus cosmologically unacceptable. Similar conclusions
can be reached for many other examples discussed in the text. In most
cases the standard matter era is replaced by a cosmic expansion with
scale factor $a\propto t^{1/2}$. We also find that $f(R)$ models
can have a strongly phantom attractor but in this case there is no
acceptable matter era. 
\end{abstract}
\maketitle

\section{Introduction}

The late-time accelerated expansion of the universe is a major challenge
to present-day cosmology (see Refs.~\cite{review,CST} for review).
A consistent picture, the concordance model, seems to emerge from
the bulk of observations probing the background evolution of the universe
as well as its inhomogeneities: Supernovae Ia \cite{SN}, Cosmic Microwave
Background anisotropies (CMB) \cite{CMB}, Large Scale Structure formation
(LSS) \cite{LSS}, baryon oscillations \cite{BAO}, weak lensing \cite{WL},
etc. If one assumes today a flat universe with a cosmological constant
$\Lambda$ and with pressureless matter, observations suggest the
following cosmological parameters $\Omega_{\Lambda,0}\approx0.7$,
$\Omega_{{\rm m},0}\approx0.3$ where $\Omega_{X,0}=\rho_{X,0}/\rho_{{\rm cr},0}$
for any component $X$, where the subscript $0$ stands for present-day
values and $\rho_{{\rm cr}}$ is the critical density of the universe.

A cosmological constant term is the simplest possibility to explain
the observational data. In fact the recent data analysis \cite{Sel}
combining the SNLS data \cite{Astier} with CMB, LSS and the Lyman-$\alpha$
forest shows that, \textit{assuming} $w_{{\rm DE}}$ is constant the
equation of state parameter of Dark Energy (DE) is found to be $w_{{\rm DE}}=-1.04\pm0.06$
and therefore consistent with a cosmological constant. However a cosmological
constant suffers from an extreme fine-tuning problem of its energy
scale if it originates from vacuum energy. For this reason several
works have explored alternative explanations, i.e., dynamical forms
of dark energy. In the absence of any compelling dynamical dark energy
model, further insight can be gained by considering general models
with constant equation of state or some fiducial parametrization,
see e.g. \cite{CP01}, and \cite{SS06} for a recent review.

The first alternative possibility to a cosmological constant is a
minimally coupled scalar field $\phi$, usually called quintessence
\cite{quin}. In analogy with inflationary scenarios, this scalar
field would be responsible for a stage of accelerated expansion while
in contrast to inflation this stage occurs in the late-time evolution
of the universe. The energy density of the scalar field should therefore
come to dominate over other components in the universe only recently.
This is the so called cosmic coincidence problem faced by most dark
energy models. In order to alleviate this problem various generalizations
have been considered, like coupled quintessence models \cite{coupled}
in which matter and dark energy scale in the same way with time during
some epochs. It is however still a challenging task to construct viable
scaling models which give rise to a matter-dominated era followed
by an accelerated scaling attractor \cite{AQTW}.

An important limitation of standard quintessence models is that they
do not allow for a phantom regime with $w_{{\rm DE}}<-1$. A phantom
regime is allowed by observations and even favored by some analysis
of the data \cite{Mel}. To achieve $w_{{\rm DE}}<-1$, the scalar
field should be endowed with a generalized kinetic term, for instance
one with a sign opposite to the canonical one \cite{phantom}. This
intriguing possibility is however plagued by quantum instabilities
\cite{Cline}. A further interesting possibility is provided by non-minimally
coupled scalar fields \cite{nonmin} and scalar-tensor cosmology \cite{stensor,Esp}.
Scalar-tensor DE models can have a consistent phantom regime and a
modified growth rate of structure \cite{BEPS00}, see also \cite{GPRS06}
for a systematic study of the low redshift structure of such theories
including a detailed analysis of the possibility to have a phantom
regime and the constraints from local gravity tests, and \cite{Peri}
for some concrete examples of this scenario. In scalar-tensor DE models
gravity is modified by an additional dynamical degree of freedom,
the scalar partner of the graviton.

Recently there has been a burst of activity dealing with so-called
modified gravity DE models (see Ref.~\cite{CST} for recent review
and references therein). In these theories one modifies the laws of
gravity whereby a late-time accelerated expansion is produced without
recourse to a DE component, a fact which renders these models very
attractive. In some models one can have in addition a phantom regime,
which might constitute an interesting feature.

The simplest family of modified gravity DE models is obtained by replacing
the Ricci scalar $R$ in the usual Hilbert-Einstein Lagrangian density
for some function $f(R)$. In the first models proposed in DE literature,
where a term $1/R$ is added to $R$ \cite{Capo,Carroll}, one typically
expects that as the universe expands the inverse curvature term will
dominate and produce the desired late-time accelerated expansion (see
Ref.~\cite{coa} for a pioneering work in the context of inflation).
However it was quickly realized that local gravity constraints would
make these models non viable \cite{Chiba} (see also Ref.~\cite{Dolgov}).
Indeed, it was shown that $f(R)$ models are formally equivalent to
scalar-tensor models with a vanishing Brans-Dicke parameter $\omega_{{\rm BD}}=0$.
Clearly such models do not pass local gravity (solar system) constraints,
in particular the post Newtonian parameter $\gamma_{{\rm PPN}}$ satisfies
$\gamma_{{\rm PPN}}=1/2$ instead of being very close to 1 as required
by observations.

However, the question of whether local gravity constraints rule out
or not $f(R)$ models does not seem to be completely settled in the
literature \cite{Faraoni}. Several papers pointed out that local
gravity constraints cannot yet rule out all possible forms of $f(R)$
theories. For instance, a model containing a particular combination
of $1/R$ and $R^{2}$ terms was suggested \cite{NO03} and claimed
by their authors to pass successfully the solar system constraints,
due to a large (infinite) effective mass needed to satisfy solar system
constraints, and \textit{also} to produce a late-time accelerated
expansion (though this latter property does not seem to have been
demonstrated in a satisfactory way). Another original approach with
negative and positive power terms was suggested recently where the
positive power term would dominate on small scales while the negative
power term dominates on large cosmic scales thereby producing the
accelerated expansion \cite{Brook} (see however \cite{nva}). See
Refs.~\cite{fRpapers} for a list of recent research in $f(R)$ dark
energy models. If $f(R)$ models are not ruled out by local gravity
constraints it is important to understand their cosmological properties.

Recently three of the present authors \cite{APT} have shown that
the large redshift behavior of $f(R)=R+\alpha R^{-n}$ models generically
lead to the {}``wrong'' expansion law: indeed, the usual matter
era preceding the late-time accelerated stage does not have the usual
$a\propto t^{2/3}$ behavior but rather $a\propto t^{1/2}$ which
would obviously make these models cosmologically unacceptable. This
intriguing and quite unexpected property of these $f(R)$ models was
overlooked in the literature. The absence of the standard matter epoch
is associated with the fact that in the Einstein frame non-relativistic
matter is strongly coupled to gravity except for the $f(R)$ theories
which have a linear dependence of $R$ (including the $\Lambda$CDM
model: $f(R)=R-\Lambda)$ \cite{APT}.

In the Einstein frame the power-law models $f(R)\propto R^{-n}$ ($n\neq-1$)
correspond to a coupled quintessence scenario with an exponential
potential of a dynamical scalar field. In this case the standard matter
era is replaced by a {}``$\phi$ matter-dominated epoch'' ($\phi$MDE)
in which the scale factor in the Einstein frame evolves as $a_{E}\propto t_{E}^{3/5}$
\cite{APT}. Transforming back to the Jordan frame, this corresponds
to a non-standard evolution $a\propto t^{1/2}$. We wish to stress
here that cosmological dynamics obtained in the Jordan frame exhibits
no difference from the one which is transformed to the Einstein frame
and transformed back to the original frame. Hence in this paper we
shall focus on the analysis in the Jordan frame without referring
to the Einstein frame.

This paper is devoted to explaining in detail our previous result
and, more importantly, to extend it to all well-behaved $f(R)$ Lagrangians.
Despite almost thirty years of work on the cosmology of $f(R)$ models,
there are in fact no general criteria in literature to gauge their
validity as alternative cosmological models (see Ref.~\cite{key-2}
for one of the earliest attempt in this direction). We find the general
conditions for a $f(R)$ theory to contain a standard matter era followed
by an accelerated attractor in a spatially flat, homogeneous and isotropic
background. The only conditions we assume throughout this paper, beside
obviously a well-behaved function $f(R)$ continuous with all its
derivatives, is that ${\rm d}f/{\rm d}R>0$, to maintain a positive
effective gravitational constant in the limit of vanishing higher-order
term. In some cases however we consider $f(R)$ models which violate
this condition in some range of $R$, but not on the actual cosmological
trajectories. The main result of this paper is that we are able to
show analytically and numerically that all $f(R)$ models with an
accelerated global attractor belong to one of four classes:

\begin{itemize}
\item Class I\,:~Models of this class possess a peculiar scale factor
behavior ($a\propto t^{1/2}$) just before the acceleration. 
\item Class II\,:~Models of this class have a matter epoch and are asymptotically
equivalent to (and hardly distinguishable from) the $\Lambda$CDM
model ($w_{{\rm eff}}=-1$). 
\item Class III\,:~Models of this class can possess an approximate matter
era but this is a transient state which is rapidly followed by the
final attractor. Technically, the eigenvalues of the matter saddle
point diverge and is very difficult to find initial conditions that
display the approximated matter epoch. 
\item Class IV$\,$: Models of this class behave in an acceptable way. They
possess an approximate standard matter epoch followed by a non-phantom
acceleration ($w_{{\rm eff}}>-1$). 
\end{itemize}
We can then summarize our findings by saying that $f(R)$ dark energy
models are either wrong (Class I), or asymptotically de-Sitter (Class
II), or strongly phantom (Class III) or, finally, standard DE (Class
IV). The second and fourth classes have some chance to be cosmologically
acceptable, but even for these cases it is not an easy task to identify
the basin of attraction of the acceptable trajectories. We fully specify
the conditions under which any given $f(R)$ model belongs to one
of the classes above and discuss analytically and numerically several
examples belonging to all classes.

An important clarification is here in order. It is clear that $f(R)$
gravity models can be perfectly viable in different contexts. The
most famous example is provided by Starobinsky's model, $f(R)=R+\alpha R^{2}$
\cite{star}, which has been the first internally consistent inflationary
model. In this model, the $R^{2}$ term produces an accelerated stage
in the early universe \emph{preceding} the usual radiation and matter
stages. A late-time acceleration in this model (after the matter dominated
stage) requires a positive cosmological constant (or some other form
of dark energy) in which case the $R^{2}$ term is no longer responsible
for the late-time acceleration.

Our paper is organized in the following way. Section II contains the
basic equations in the Jordan frame and introduces autonomous equations
which are applicable to any forms of $f(R)$. In Sec.~III we derive
fixed points together with their stabilities and present the conditions
for viable $f(R)$ DE models. In Sec.~VI we classify $f(R)$ DE models
into four classes depending upon the cosmological evolution which
gives the late-time acceleration. In Sec.~V we shall analytically
show the cosmological viability for some of the $f(R)$ models by
using the conditions found in Sec.~III. Section VI is devoted to
a numerical analysis for a number of $f(R)$ models to confirm the
analytical results presented in the previous section. Finally we summarize
our results in Section VII. We will always work in the Jordan frame,
in order to show the properties of $f(R)$ models in the most direct
way, without the need to convert back from the Einstein frame.

\section{$f(R)$ dark energy models}

\subsection{Definitions and equations}

In this section we derive all basic equations in the Jordan frame
(JF), the frame in which observations are performed. We will further
define all fundamental quantities characterizing our system, in particular
the equation of state of our system. Actually, as we will see below
this is a subtle issue and we have to \textit{define} what is meant
by the Dark Energy (DE) equation of state.

We concentrate on spatially flat Friedman-Lema\^{\i}tre-Robertson-Walker
(FLRW) universes with a time-dependent scale factor $a(t)$ and a
metric \begin{equation}
\rd s^{2}=-\rd t^{2}+a^{2}(t)\,\rd{\bf x}^{2}\,.\end{equation}
 For this metric the Ricci scalar $R$ is given by \begin{equation}
R=6\left(2H^{2}+\dot{H}\right)\,,\label{R}\end{equation}
 where $H\equiv\dot{a}/a$ is the Hubble rate and a dot stands for
a derivative with respect to $t$.

We start with the following action in the JF \begin{equation}
S=\int{\rm d}^{4}x\sqrt{-g}\left[\frac{1}{2\kappa^{2}}f(R)
+{\mathcal{L}}_{{\rm rad}}+{\mathcal{L}}_{{\rm m}}\right]\,,\end{equation}
 where $\kappa^{2}=8\pi G$ while $G$ is a bare gravitational constant,
$f(R)$ is some arbitrary function of the Ricci scalar $R$, and ${\mathcal{L}}_{{\rm m}}$
and ${\mathcal{L}}_{{\rm rad}}$ are the Lagrangian densities of dust-like
matter and radiation respectively. 
Note that $G$ is typically {\emph not} Newton's gravitational constant 
measured in the attraction between two test masses in Cavendish-type 
experiments (see e.g. \cite{Esp}).
Then the following equations are obtained \cite{Jaichan}
\begin{eqnarray}
3FH^{2} & = & \kappa^{2}\,(\rho_{{\rm m}}+\rho_{{\rm rad}})+\frac{1}{2}(FR-f)-3H\dot{F}\,,\label{E1}\\
-2F\dot{H} & = & \kappa^{2}\left(\rho_{{\rm m}}+\frac{4}{3}\rho_{{\rm rad}}\right)+\ddot{F}-H\dot{F}\,,\label{E2}\end{eqnarray}
 where \begin{equation}
F\equiv\frac{\rd f}{\rd R}\,.\end{equation}
 In standard Einstein gravity ($f=R$) one has $F=1$. In what follows
we shall consider the positive-definite forms of $F$ to avoid a singularity
at $F=0$. The densities $\rho_{{\rm m}}$ and $\rho_{{\rm rad}}$
satisfy the usual conservation equations \begin{eqnarray}
 &  & \dot{\rho}_{{\rm m}}+3H\rho_{{\rm m}}=0\,,\label{E3}\\
 &  & \dot{\rho}_{{\rm rad}}+4H\rho_{{\rm rad}}=0\,.\label{E4}\end{eqnarray}

We note that Eqs.~(\ref{E1}) and (\ref{E2}) are similar to those
obtained for scalar-tensor gravity \cite{BEPS00} with a vanishing
Brans-Dicke parameter $\omega_{{\rm BD}}=0$ and a specific potential
$U=(FR-f)/2$. Note that in scalar-tensor gravity we have $FR=L$
so that this term vanishes, while Eq.~(\ref{E2}) is similar except
for the fact that a kinematic term of the scalar field is absent.
Hence we can define the DE equation of state in a way similar to that
in scalar-tensor theories of gravity (see e.g., \cite{GPRS06,T02}).
With a straightforward redefinition of the quantities, we rewrite
Eqs.~(\ref{E1}) and (\ref{E2}) as follows \begin{eqnarray}
3F_{0}H^{2} & = & \kappa^{2}(\rho_{{\rm DE}}+\rho_{{\rm m}}+\rho_{{\rm rad}})\,,\label{E10}\\
-2F_{0}\dot{H} & = & \kappa^{2}\left(\rho_{{\rm m}}+\frac{4}{3}\rho_{{\rm rad}}+\rho_{{\rm DE}}+p_{{\rm DE}}\right)\,.\label{E2a}\end{eqnarray}
 We then have the following equalities \begin{eqnarray}
\kappa^{2}\,\rho_{{\rm DE}} & = & \frac{1}{2}(FR-f)-3H\dot{F}+3H^{2}(F_{0}-F)\,,\label{rhoDE}\\
\kappa^{2}\, p_{{\rm DE}} & = & \ddot{F}+2H\dot{F}-\frac{1}{2}(FR-f)-(2\dot{H}+3H^{2})(F_{0}-F)\,.\end{eqnarray}
 The energy density $\rho_{{\rm DE}}$ and the pressure density $p_{{\rm DE}}$
of DE defined in this way satisfy the usual conservation equation
\begin{equation}
{\dot{\rho}_{{\rm DE}}}=-3H(\rho_{{\rm DE}}+p_{{\rm DE}})\,.\end{equation}
 Hence the equation of state parameter $w_{{\rm DE}}$ defined through
\begin{equation}
w_{{\rm DE}}\equiv\frac{p_{{\rm DE}}}{\rho_{{\rm DE}}}=-1+\frac{2\ddot{F}-2H\dot{F}-4\dot{H}(F_{0}-F)}{(FR-f)-6H\dot{F}+6H^{2}(F_{0}-F)}\,,\end{equation}
 acquires its usual physical meaning, in particular the time evolution
of the DE sector is given by \begin{equation}
\frac{\rho_{{\rm DE}}(z)}{\rho_{{\rm DE,0}}}=\exp\left[3\int_{0}^{z}{\rm d}z'~\frac{1+w_{{\rm DE}}(z')}{1+z'}\right]\,,\end{equation}
 where $z\equiv a_{0}/a-1$. Note that the subscript {}``0'' stands
for present values. It is $\rho_{{\rm DE}}$, as defined in Eq.~(\ref{rhoDE})
which is the quantity extracted from the observations and $w_{{\rm DE}}$
the corresponding DE equation of state parameter for which specific
parametrizations are used.

Looking at Eqs.~(\ref{E10}) and (\ref{E20}), one could introduce
the cosmological parameters $\tilde{\Omega}_{X}=\kappa^{2}\rho_{X}/(3F_{0}H^{2})$
\cite{GPRS06,T02}. However here it turns out to be more convenient
to work with the density parameters \begin{equation}
\Omega_{X}\equiv~\frac{\kappa^{2}\rho_{X}}{3FH^{2}}\,,\label{Om}\end{equation}
 where $X={\rm m},{\rm rad}$ or ${\rm DE}$. The quantity $w_{{\rm DE}}$
can further be obtained directly from the observations \begin{eqnarray}
w_{{\rm {DE}}} & = & \frac{(1+z)\,\rd h^{2}/\rd z-3h^{2}-\Omega_{{\rm rad,0}}(1+z)^{4}}{3[h^{2}-\Omega_{{\rm m},0}(1+z)^{3}-\Omega_{{\rm rad},0}(1+z)^{4}]}\,,\label{wDEz}\end{eqnarray}
 where $h\equiv H/H_{0}$. In the low-redshift region where the contribution
of the radiation is negligible, we have \begin{eqnarray}
w_{{\rm {DE}}} & = & \frac{(1+z)\,\rd h^{2}/\rd z-3h^{2}}{3[h^{2}-\Omega_{{\rm m},0}(1+z)^{3}]}\,,\qquad z\ll z_{{\rm eq}}\,,\label{wDEzz}\end{eqnarray}
 where $z_{{\rm eq}}$ is the redshift at which dust and radiation
have equal energy densities. Equation (\ref{wDEz}) can be extended
for spatially non-flat universes \cite{PR05} but we restrict ourselves
to spatially flat universes. We also define the effective equation
of state \begin{equation}
w_{{\rm eff}}=-1-\frac{2\dot{H}}{3H^{2}}\,.\label{weff}\end{equation}
 Note that the following equality holds \begin{equation}
w_{{\rm eff}}=\tilde{\Omega}_{{\rm DE}}~w_{{\rm DE}}+\frac{1}{3}\tilde{\Omega}_{{\rm rad}}\,,\label{weff0}\end{equation}
 if we define $\tilde{\Omega}_{X}=\kappa^{2}\rho_{X}/(3F_{0}H^{2})$.

\subsection{Autonomous equations}

For a general $f(R)$ model it will be convenient to introduce the
following (dimensionless) variables \begin{eqnarray}
x_{1} & = & -\frac{\dot{F}}{HF}\,,\label{x1}\\
x_{2} & = & -\frac{f}{6FH^{2}}\,,\label{x2}\\
x_{3} & = & \frac{R}{6H^{2}}=\frac{\dot{H}}{H^{2}}+2\,,\label{x3}\\
x_{4} & = & ~\frac{\kappa^{2}\rho_{{\rm rad}}}{3FH^{2}}\,.\label{x4}\end{eqnarray}
 From Eq.~(\ref{E1}) we have the algebraic identity \begin{equation}
\Omega_{\m}\equiv\frac{\kappa^{2}\rho_{\m}}{3FH^{2}}=1-x_{1}-x_{2}-x_{3}-x_{4}\,.\label{Omem}\end{equation}
 It is then straightforward to obtain the following equations of motion
\begin{eqnarray}
\frac{\rd x_{1}}{\rd N} & = & -1-x_{3}-3x_{2}+x_{1}^{2}-x_{1}x_{3}+x_{4}~,\label{N1}\\
\frac{\rd x_{2}}{\rd N} & = & \frac{x_{1}x_{3}}{m}-x_{2}(2x_{3}-4-x_{1})~,\label{N2}\\
\frac{\rd x_{3}}{\rd N} & = & -\frac{x_{1}x_{3}}{m}-2x_{3}(x_{3}-2)~,\label{N3}\\
\frac{\rd x_{4}}{\rd N} & = & -2x_{3}x_{4}+x_{1}\, x_{4}\,,\label{N4}\end{eqnarray}
 where $N$ stands for $\ln a$ and \begin{eqnarray}
m & \equiv & \frac{\rd\log F}{\rd\log R}=\frac{Rf_{,RR}}{f_{,R}}\,,\label{mdef}\\
r & \equiv & -\frac{\rd\log f}{\rd\log R}=-\frac{Rf_{,R}}{f}=\frac{x_{3}}{x_{2}}\,,\label{ldef}\end{eqnarray}
 where $f_{,R}\equiv{\rm d}f/{\rm d}R$ and $f_{,RR}\equiv{\rm d}^{2}f/{\rm d}R^{2}$.
Deriving $R$ as a function of $x_{3}/x_{2}$ from Eq.~(\ref{ldef}),
one can express $m$ as a function of $x_{3}/x_{2}$ and obtain the
function $m(r)$. For the power-law model with $f(R)=\alpha R^{-n}$
the variable $m$ is a constant ($m=-n-1$) with $r=n=x_{3}/x_{2}$.
In this case the system reduces to a 3-dimensional one with variables
$x_{1}$, $x_{2}$ and $x_{4}$. However for general $f(R)$ gravity
models the variable $m$ depends upon $r$.

We also make use of these expressions: \begin{eqnarray}
w_{{\rm eff}} & = & -\frac{1}{3}(2x_{3}-1)\,,\label{weffx}\\
w_{{\rm DE}} & = & \frac{1}{3}\frac{1-x_{4}y-2x_{3}}{1-y(1-x_{1}-x_{2}-x_{3})}\,,\label{wDEx}\end{eqnarray}
 where $y=F/F_{0}$.

\section{Cosmological dynamics of $f(R)$ gravity models}

In this section we derive the analytical properties of the phase space.

\subsection{Critical points and stability for a general $f(R)$}

In the absence of radiation ($x_{4}=0$) the critical points for the
system (\ref{N1})-(\ref{N3}) for any $m(r)$ are \begin{eqnarray}
 &  & P_{1}:(x_{1},x_{2},x_{3})=(0,-1,2),\quad\Omega_{\m}=0,\quad w_{{\rm eff}}=-1\,,\\
 &  & P_{2}:(x_{1},x_{2},x_{3})=(-1,0,0),\quad\Omega_{\m}=2,\quad w_{{\rm eff}}=1/3\,,\\
 &  & P_{3}:(x_{1},x_{2},x_{3})=(1,0,0),\quad\Omega_{\m}=0,\quad w_{{\rm eff}}=1/3\,,\\
 &  & P_{4}:(x_{1},x_{2},x_{3})=(-4,5,0),\quad\Omega_{\m}=0,\quad w_{{\rm eff}}=1/3\,,\\
 &  & P_{5}:(x_{1},x_{2},x_{3})=\left(\frac{3m}{1+m},-\frac{1+4m}{2(1+m)^{2}},\frac{1+4m}{2(1+m)}\right),\quad\Omega_{\m}=1-\frac{m(7+10m)}{2(1+m)^{2}},\quad w_{{\rm eff}}=-\frac{m}{1+m}\,,\\
 &  & P_{6}:(x_{1},x_{2},x_{3})=\left(\frac{2(1-m)}{1+2m},\frac{1-4m}{m(1+2m)},-\frac{(1-4m)(1+m)}{m(1+2m)}\right),\quad\Omega_{\m}=0,\quad w_{{\rm eff}}=\frac{2-5m-6m^{2}}{3m(1+2m)}\,,\end{eqnarray}
 where here $\Omega_{\m}=1-x_{1}-x_{2}-x_{3}$.

The points $P_{5}$ and $P_{6}$ satisfy the equation $x_{3}=-(m(r)+1)x_{2}$,
i.e., \begin{equation}
m(r)=-r-1\,.\label{P5P6}\end{equation}
 When $m(r)$ is not a constant, one must solve this equation. For
each root $r_{i}$ one gets a point of type $P_{5}$ or $P_{6}$ with
$m=m(r_{i})$. For instance, the $f(R)=R+\alpha R^{-n}$ model corresponds
to $m(r)=-n(1+r)/r$ as we will see later, which then gives $r_{1,2}=-1,n$
and $m_{1,2}=0,-1-n$. If we assume that $m={\rm constant}$ then
the condition $x_{3}=-(m+1)x_{2}$ must hold from Eqs.~(\ref{mdef})
and (\ref{ldef}). Hence for $m={\rm constant}$ the points $P_{2,3,5,6}$
always exist, while $P_{1}$ and $P_{4}$ are present for $m=1$ and
$m=-1$ respectively. The solutions which give the exact equation
of state of a matter era ($w_{{\rm eff}}=0$, i.e., $a\propto t^{2/3}$
or $x_{3}=1/2$) exist only for $m=0$ ($P_{5}$) or for $m=-(5\pm\sqrt{73})/12$
($P_{6}$) \cite{Capo06}. However the latter case corresponds to
$\Omega_{\m}=0$, so this does not give a standard matter era dominated
by a non-relativistic fluid \cite{APT2}.

If $m(r)$ is not constant then there can be any number of distinct
solutions, although only $P_{1}$ and those originating from $P_{5,6}$
can be accelerated and only $P_{2}$ and $P_{5}$ might give rise
to matter eras. However $P_{2}$ corresponds to $w_{{\rm eff}}=1/3$
and therefore is ruled out as a correct matter era: this is in fact
the $a\propto t^{1/2}$ behavior discussed in Ref.~\cite{APT} (and
denoted as $\phi$MDE since it is in fact a field-matter dominated
epoch in the Einstein frame). On the contrary, $P_{5}$ resembles
a standard matter era, but only for $m$ close to 0. Hence a {}``good''
cosmology would be given by any trajectory passing near $P_{5}$ with
$m$ close to 0 and landing on an accelerated attractor. Any other
behavior would not be consistent with observations.

It is important to realize that the surface $x_{2},x_{3}$ for which
$m(r)=-r-1$ is a subspace of the system (\ref{N1}-\ref{N4}) and
therefore it cannot be crossed. This can be seen by using the definition
of $r$ and $m$ to derive the following equation for $r$: \begin{equation}
\frac{{\rm d}r}{{\rm d}N}=r(1+m+r)\frac{\dot{R}}{HR}\,,\label{eq:critline}\end{equation}
 which shows explicitly that $m=-r-1$ implies ${\rm d}r/{\rm d}N=0$
as long as $\dot{R}/HR$ does not diverge. This means that the evolution
of the system along the $m(r)$ line stops at the roots of the equation
$m=-r-1$ so that every cosmological trajectory is trapped between
successive roots.

In what follows we shall consider the properties of each fixed point
in turn. We define $m_{i}\equiv m(P_{i})$ and will always assume
a general $m=m(r)$.

\begin{itemize}
\item (1) $P_{1}$: de-Sitter point

Since $w_{{\rm eff}}=-1$ the point $P_{1}$ corresponds to de-Sitter
solutions ($\dot{H}=0$) and has eigenvalues \begin{equation}
-3,~~~-\frac{3}{2}\pm\frac{\sqrt{25-16/m_{1}}}{2}\,,\end{equation}
 where $m_{1}=m(r=-2)$. Hence $P_{1}$ is stable when $0<m_{1}\le1$
and a saddle point otherwise. Then the condition for the stability
of the de-Sitter point is given by \begin{equation}
0\le m(r=-2)\le1\,.\label{m1con}\end{equation}

\item (2) $P_{2}$: $\phi$MDE

Point $P_{2}$ is characterized by a {}``kinetic'' epoch in which
matter and field co-exist with constant energy fractions. We denote
it as a $\phi$-matter dominated epoch ($\phi$MDE) following Ref.~\cite{coupled}.
The eigenvalues are given by \begin{eqnarray}
-2,~~\frac{1}{2}\left[7+\frac{1}{m_{2}}-\frac{m_{2}'}{m_{2}^{2}}r(1+r)\pm\sqrt{\left\{ 7+\frac{1}{m_{2}}-\frac{m_{2}'}{m_{2}^{2}}r(1+r)\right\} ^{2}-4\left\{ 12+\frac{3}{m_{2}}-\frac{m_{2}'}{m_{2}^{2}}r(3+4r)\right\} }\right]\,,\end{eqnarray}
 where a prime represents a derivative with respect to $r$. Hence
$P_{2}$ is either a saddle or a stable node. If $m(r)$ is a constant
the eigenvalues reduce to $-2,3,4+1/m_{2}$, in which case $P_{2}$
is a saddle point. Note that it is stable on the subspace $x_{3}=rx_{2}$
for $-1/4<m<0$. However, from Eqs.~(\ref{N2}) and (\ref{N3}),
one must ensure that the term $x_{3}/m_{2}$ vanishes. Then the necessary
and sufficient condition for the existence of the point $P_{2}$ is
expressed simply by \begin{equation}
\lim_{x_{2,3}\to0}\frac{x_{3}}{m_{2}}=0\,,\label{limitm}\end{equation}
 which amounts to \begin{equation}
\frac{f_{,R}}{H^{2}f_{,RR}}\to0\,,\end{equation}
 for $R/H^{2}\to0$ and $f/f_{,R}H^{2}\to0$. This applies immediately
to several models, like e.g., $f=\log R,R^{n},R+\alpha R^{n}$ and
in general for any well-behaved $f(R)$, i.e., for all the functions
that satisfy the condition of application of de l'Hopital rule. This
shows that the {}``wrong'' matter era is indeed generic to the $f(R)$
models.

\item (3) $P_{3}$: Purely kinetic point

This also corresponds to a {}``kinetic'' epoch, but it is different
from the point $P_{2}$ in the sense that the energy fraction of the
matter vanishes. Point $P_{3}$ can be regarded as the special case
of the point $P_{6}$ by setting $m=1/4$. The eigenvalues are given
by \begin{eqnarray}
2,~~\frac{1}{2}\left[9-\frac{1}{m_{3}}+\frac{m_{3}'}{m_{3}^{2}}r(1+r)\pm\sqrt{\left\{ 9-\frac{1}{m_{3}}+\frac{m_{3}'}{m_{3}^{2}}r(1+r)\right\} ^{2}-4\left\{ 20-\frac{5}{m_{3}}+\frac{m_{3}'}{m_{3}^{2}}r(5+4r)\right\} }\right]\,,\end{eqnarray}
 which means that $P_{3}$ is either a saddle or an unstable node.
If $m(r)$ is a constant the eigenvalues reduce to $2,5,4-1/m_{3}$.
In this case $P_{3}$ is unstable for $m_{3}<0$ and $m_{3}>1/4$
and a saddle otherwise.

\item (4) $P_{4}$

This point has a similar property to $P_{3}$ because both ${\Omega}_{\m}$
and $w_{{\rm eff}}$ are the same as those of $P_{3}$. It is regarded
as the special case of the point $P_{6}$ by setting $m=-1$. Point
$P_{4}$ has eigenvalues \begin{equation}
-5,~~-3,~~4\left(1+1/m_{4}\right)\,.\end{equation}
 Hence it is stable for $-1<m_{4}<0$ and a saddle otherwise. Neither
of $P_{3}$ and $P_{4}$ can be used for the matter-dominated epoch
nor for the accelerated epoch.

\item (5) $P_{5}$: Scaling solutions

Point $P_{5}$ corresponds to scaling solutions which give the constant
ratio $\Omega_{\m}/\Omega_{{\rm DE}}$. In the limit $m_{5}\to0$,
it actually represents a standard matter era with $a\propto t^{2/3}$
and $\Omega_{\m}=1$. Hence the necessary condition for $P_{5}$ to
exist as an exact standard matter era is given by \begin{eqnarray}
m(r=-1)=0\,.\label{m5con}\end{eqnarray}
 The eigenvalues of $P_{5}$ are given by \begin{eqnarray}
3(1+m_{5}'),~~\frac{-3m_{5}\pm\sqrt{m_{5}(256m_{5}^{3}+160m_{5}^{2}-31m_{5}-16)}}{4m_{5}(m_{5}+1)}\,.\label{P5eig}\end{eqnarray}
 In the limit $|m_{5}|\ll1$ the eigenvalues approximately reduce
to \begin{equation}
3(1+m_{5}'),~~-\frac{3}{4}\pm\sqrt{-\frac{1}{m_{5}}}\,.\label{eig}\end{equation}

The models with $m_{5}=Rf_{,RR}/f_{,R}<0$ exhibit the divergence
of the eigenvalues as $m_{5}\to-0$, in which case the system cannot
remain for a long time around the point $P_{5}$. For example the
models $f(R)=R-\alpha/R^{n}$ with $n>0$ and $\alpha>0$ \cite{Capo,Carroll}
fall into this category. An approximate matter era exists also if
instead $m_{5}$ is negative and non-zero but then the eigenvalues
are large and it is difficult to find initial conditions that remain
close to it for a long time. We shall present such an example in a
later section. Therefore generally speaking models with $m_{5}<0$
are not acceptable, except at most for a very narrow range of initial
conditions. On the other hand, if $0<m_{5}<0.327$ the latter two
eigenvalues in Eq.~(\ref{P5eig}) are complex with negative real
parts. Then, provided that $m_{5}'>-1$, the point $P_{5}$ can be
a saddle point with a damped oscillation. Hence in principle the universe
can evolve toward the point $P_{5}$ and then leave for the late-time
acceleration. Note that the point $P_{2}$ is also generally a saddle
point except for some specific cases in which it is stable. Which
trajectory ($P_{2}$ or $P_{5}$) is chosen depends upon initial conditions,
so a numerical analysis is necessary.

Note that from the relation (\ref{P5P6}) the condition $m(r)>0$
is equivalent to $r<-1$. Hence the criterion for the existence of
a saddle matter epoch with a damped oscillation is given by \begin{eqnarray}
m(r\leq-1)>0\,,\quad m'(r\leq-1)>-1\,.\label{fRR}\end{eqnarray}

Note that we also require the condition (\ref{m5con}). In order to
realize an accelerated stage after the matter era, additional conditions
are necessary as we will discuss below. Finally, we remark that a
special case occurs if $m={\rm const}$. This corresponds to $f(R)=-\Lambda+\alpha R^{-n}$.
In this case the system contains a two-dimensional subspace $x_{3}=-(m+1)x_{2}=nx_{2}$
and on this subspace the stability of the latter two eigenvalues in
Eq.~(\ref{P5eig}) is sufficient to ensure the stability. Working
with the $(x_{1},x_{2},x_{3})$ phase space the trajectories that
start with $x_{3}=nx_{2}$, which implies $\Lambda=0$, remain on
the subspace. Then the point is stable in the range $0<m_{5}<0.327$
. For $\Lambda\not=0$, the trajectories start off the subspace and
follow the same criteria of stability as for the $m\not=$const. case.
So there exists a standard saddle matter era for $f(R)=-\Lambda+\alpha R^{1+\epsilon}$
with $\epsilon$ small and positive.

\item (6) $P_{6}$: Curvature-dominated point

This corresponds to the curvature-dominated point whose effective
equation of state depends upon the value $m$. It satisfies the condition
for acceleration ($w_{{\rm eff}}<-1/3$) when $m_{6}<-(1+\sqrt{3})/2$,
$-1/2<m_{6}<0$ and $m_{6}>(\sqrt{3}-1)/2$. In Fig.~\ref{fig:p6}
we show the behavior of $w_{{\rm eff}}$ as a function of $m$. The
eigenvalues are given by \begin{equation}
-4+\frac{1}{m_{6}},~~\frac{2-3m_{6}-8m_{6}^{2}}{m_{6}(1+2m_{6})},~~-\frac{2(m_{6}^{2}-1)(1+m'_{6})}{m_{6}(1+2m_{6})}\,.\label{eigenP6}\end{equation}
 Hence the stability of $P_{6}$ depends on both $m_{6}$ and $m'_{6}$.
In the limit $m_{6}\to\pm\infty$ we have $P_{6}\to(-1,0,2)$ with
a de-Sitter equation of state ($w_{{\rm eff}}\to-1$). This point
is stable provided that $m_{6}'>-1$. $P_{6}$ is also a de-Sitter
point for $m_{6}=1$, which coincides with $P_{1}$ and is marginally
stable. Since $r=-2$ in this case, this point is characterized by
\begin{equation}
m(r=-2)\to1\,.\end{equation}
 It is instructive to see this property in the Einstein frame, i.e.
performing a conformal transformation of the system \cite{APT}. Then
one obtains a scalar field with a potential $V=(FR-f)/|F|^{2}$. This
shows that the condition $m(r=-2)=1$ corresponds to $V_{,R}=0$,
i.e., the condition for the existence of a potential minimum.

The point $P_{6}$ is both stable and accelerated in four distinct
ranges. \\

{[}I] $m_{6}'>-1$

When $m_{6}'>-1$, $P_{6}$ is stable and accelerated in the following
three regions:

\begin{itemize}
\item (A) $m_{6}<-(1+\sqrt{3})/2$: $P_{6}$ is accelerated but not a phantom,
i.e., $w_{{\rm eff}}>-1$. One has $w_{{\rm eff}}\to-1$ in the limit
$m_{6}\to-\infty$. 
\item (B) $-1/2<m_{6}<0$: $P_{6}$ is strongly phantom with $w_{{\rm eff}}<-7.6$. 
\item (C) $m_{6}\ge1$: $P_{6}$ is slightly phantom with $-1.07<w_{{\rm eff}}\le-1$.
One has $w_{{\rm eff}}\to-1$ in the limit $m_{6}\to+\infty$ and
$m_{6}\to1$. \\

\end{itemize}
{[}II] $m_{6}'<-1$

When $m_{6}'<-1$, the point $P_{6}$ is stable and accelerated in
the region

\begin{itemize}
\item (D) $(\sqrt{3}-1)/2<m_{6}<1$: here $P_{6}$ is a non-phantom, $w_{{\rm eff}}>-1$. 
\end{itemize}
Therefore from this we derive the first general conclusion concerning
$f(R)$ models: the asymptotic acceleration cannot have an equation
of state in the range $-7.6<w_{{\rm eff}}<-1.07$.

\end{itemize}
\begin{figure}
\includegraphics{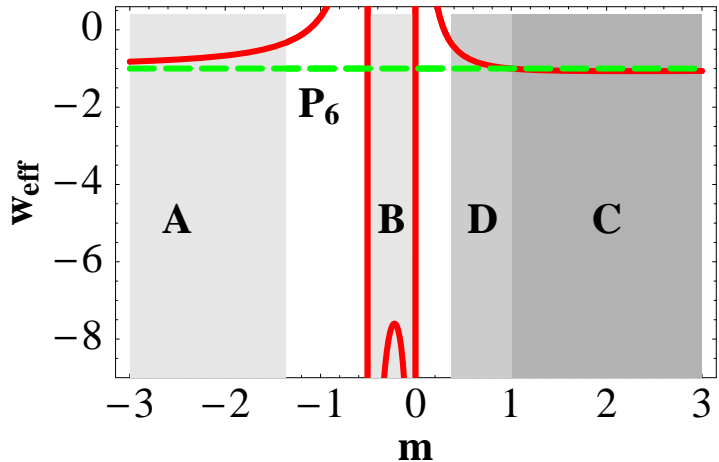}

\caption{\label{fig:p6} The effective equation of state $w_{{\rm eff}}$
for $P_{6}$ as a function of $m$. The point is stable and accelerated
in the grayed regions. In the region (A) $m<-(\sqrt{3}+1)/2$ the
point is always a non-phantom ($w_{{\rm eff}}>-1$); in the region
(B) $-1/2<m<0$ it is strongly a phantom ($w_{{\rm eff}}<-7.6$);
in the region (C) $m\ge1$ it is slightly phantom ($-1.07<w_{{\rm eff}}\le-1$)
and in the region (D) is a non-phantom ($w_{{\rm eff}}>-1$). In all
the other regions $P_{6}$ is either decelerated or unstable. Notice
the gap between $w_{{\rm eff}}=-1.07$ and $-7.6$.}
\end{figure}

If one considers radiation in addition to $x_{1,2,3}$, then all the
points $P_{1-6}$ remain the same (with $x_{4}=0$) and one obtains
two additional points \begin{eqnarray}
\bullet &  & (7)~~P_{7}:(x_{1},x_{2},x_{3},x_{4})=(0,0,0,1),\quad\Omega_{\m}=0,\quad w_{{\rm eff}}=1/3\,,\\
\bullet &  & (8)~~P_{8}:(x_{1},x_{2},x_{3},x_{4})=\left(\frac{4m}{1+m},-\frac{2m}{(1+m)^{2}},\frac{2m}{1+m},\frac{1-2m-5m^{2}}{(1+m)^{2}}\right),\quad\Omega_{\m}=0\,,\quad w_{{\rm eff}}=\frac{1-3m}{3+3m}\,.\end{eqnarray}

We see that $P_{7}$ is a standard radiation point. When $m(r)$ is
a constant the eigenvalues of $P_{7}$ are given by $1,4,4,-1$, which
means that $P_{7}$ is a saddle in this case. The point $P_{8}$ is
a new radiation era (we call it {}``$\phi$ -radiation dominated
epoch'') which contains non-zero dark energy. Since the effective
equation of state is constrained by nucleosynthesis to be close to
$1/3$, $P_{8}$ is acceptable as a radiation epoch only for $m_{8}$
close to 0.

The eigenvalues of $P_{8}$ are given by \begin{eqnarray}
1,~~4(1+m_{8}'),~~\frac{m_{8}-1\pm\sqrt{81m_{8}^{2}+30m_{8}-15}}{2(m_{8}+1)}\,.\end{eqnarray}
 In the limit $m_{8}\to0$ the last two are complex with negative
real parts, which then shows that $P_{8}$ is a saddle around the
radiation point. Hence the solutions eventually repel away from the
radiation era and are followed by one of the fixed points given above.
Unlike the matter point $P_{5}$ there are no singularities for the
eigenvalues of $P_{8}$ in the limit $m_{8}\to0$. We also note that
$P_{8}$ is on the line $m=-r-1$ as in the case of the matter point
$P_{5}$. If the condition for the existence of the matter point $P_{5}$
is satisfied (i.e., $m\approx0$ and $r\approx-1$), there exists
a radiation point $P_{8}$ in the same region. Then a viable cosmological
trajectory starts around the radiation point $P_{8}$ with $m\approx0$
and then connects to the matter point $P_{5}$ with $m\approx0$.
Finally the solutions approach either of the accelerated points mentioned
above.

\section{Four classes of Models}

For a cosmological model to work, it has to possess a matter dominated
epoch followed by an accelerated expansion. In our scenarios this
would be a stable acceleration (late-time attractor). We require that
the matter era is long enough to allow for structure formation and
that an effective equation of state is close to $w_{{\rm eff}}=0$
in order to match the observations of the diameter distance of acoustic
peaks of CMB anisotropies, i.e., it has to expand as $a\sim t^{2/3}$.
Now we study the conditions under which these requirements are met.

Let us recall again that $P_{2}$ exists as a saddle or a stable node.
Then the $\phi$MDE is always present provided that the condition
(\ref{limitm}) is satisfied and only by a choice of initial conditions
one can escape it. Hence below we examine the cases in which initial
conditions exist such that the standard matter era $P_{5}$ for $|m|\ll1$
is also a saddle. When this is possible, a numerical analysis is necessary
to ascertain the basin of attraction of $P_{2}$ and $P_{5}$. In
particular, it is necessary to see whether initial conditions that
allow for a radiation epoch lead to $P_{2}$ or $P_{5}$. Then, if
$P_{5}$ exists and is a saddle, we examine the conditions for a late-time
accelerated attractor.

\subsection{Transition from the matter point $P_{5}$ to an accelerated point
$P_{6}$ or $P_{1}$}

The only point which allows for a standard matter era is $P_{5}$
when $m(-1)\to0$, so this is the first condition for a theory to
be acceptable. If $m(-1)$ is non-vanishing the matter epoch can be
characterized by $a\sim t^{2(1+m)/3}$, which is still acceptable
if $|m|\ll1$. So from now on when we write $m(-1)\to0$ we always
mean $|m(-1)|\ll1$. The corresponding point $P_{5}$ with $|m_{5}|\ll1$
will be denoted as $P_{5}^{(0)}$. In the general case, Eq.~(\ref{P5P6})
has several roots $r_{a,b..}$ and therefore $m_{a,b..}$ and correspondingly
there will be several points $P_{5(a,b,...)},P_{6(a,b,..)}$ . Let
us call the line $m=-r-1$ on the $(r,m)$ plane the \emph{critical
line}, since the points $P_{5}$ and $P_{6}$ lie on this line. From
the matter epoch $P_{5}^{(0)}$at $(r,m)=(-1,0)$, the trajectories
can reach an acceleration point at either $P_{1}$ or one of the points
$P_{5}$ (beside $P_{5}^{(0)}$) or $P_{6}$, the only points that
can be accelerated. The point $P_{1}$ is stable and accelerated only
for $0<m_{1}\le1$. The point $P_{5}$ corresponds to an accelerated
solution for $m_{5}>1/2$ and $m_{5}<-1$; however, it can be shown
that it is not stable (saddle or unstable node) in both regions. Therefore
we only need to study the transition from the matter point $P_{5}^{(0)}$
to the accelerated point $P_{6}$. Generally speaking, a $f(R)$ model
is cosmologically viable if one of the transition $P_{5}^{(0)}\to P_{6}$
or $P_{5}^{(0)}\to P_{1}$ is possible.

The point $P_{5}^{(0)}:(r,m)=(-1,0)$ can be approached from the positive
$m$ side or from the negative one. In the first case, two eigenvalues
are complex while the real part of the eigenvalues remains finite
and negative. Then the trajectory exhibits a damped oscillation around
the matter point, before leaving for the acceleration. In the second
case, the eigenvalues are real and diverge for $m\to-0$. Then the
matter era is very short and it is very difficult to find initial
conditions that lead to a successful cosmology. The pure power-law
model $f(R)=\alpha R^{-n}$ is a special case because then $P_{5}^{(0)}$
is actually stable for $m=-1-n$ small and positive, so it is not
possible to reach the acceleration at $P_{6}$ {[}note that in this
case the system is two-dimensional with the latter two eigenvalues
in Eq.~(\ref{eig})]. For the model $f(R)=-\Lambda+\alpha R^{1+\epsilon}$
with $\epsilon$ small and positive, the transition from $P_{5}^{(0)}$
to $P_{6}$ is instead possible and these models are cosmologically
acceptable. This shows that a $\Lambda$CDM cosmology is recovered
for this model in the limit $\epsilon\to+0$ but not $\epsilon\to-0$.

As we have seen in the previous section, the point $P_{6}$ is stable
and accelerated in four distinct regions: (A) $m_{6}<-(1+\sqrt{3})/2$,
(B) $-1/2<m_{6}<0$, (C) $m_{6}\ge1$ (all these are stable if $m_{6}'>-1$);
and finally, if $m_{6}'<-1$, (D) $(\sqrt{3}-1)/2<m_{6}<1$. In the
regions (A) and (D) the point $P_{6}$ leads to a non-phantom acceleration
with $w_{{\rm eff}}>-1$, whereas the region (B) corresponds to a
strongly phantom ($w_{{\rm eff}}<-7.6$) and the region (C) to a slightly
phantom ($-1.07<w_{{\rm eff}}\le-1$). In what follows we shall discuss
each case separately.

\subsubsection{From $P_{5}$ $(m_{5}'>-1,m>0)$ to $P_{1}$ or to $P_{6}$ $(m_{6}'>-1)$
in the regions (A), (B), (C)}

In the positive $m$ region the matter point $P_{5}^{(0)}$ is a saddle
for $m_{5}'>-1$. We require the condition $m_{6}'>-1$ for the stability
of the point $P_{6}$ in the regions (A), (B) and (C). Let us then
assume that beside the root at $m\approx+0$ there are three roots
which exist in the regions (A), (B), (C), i.e. $m_{6a}$, $m_{6b}$,
$m_{6c}$, respectively. A good cosmology goes from a saddle $P_{5}^{(0)}$
to a stable acceleration, either $P_{6a},P_{6b},P_{6c}$ or $P_{1}$.
Now $P_{5}^{(0)}$ is a saddle if $m_{5}'>-1$, while $P_{6}$ is
stable if $m_{6}'>-1$. This shows that the curve $m(r)$ must intersect
the critical point line $m=-r-1$ with a derivative $m_{5,6}'>-1$.
If the intersection occurs with a derivative $m_{5,6}'<-1$, the cosmological
model is unacceptable, either because the matter era is stable or
because the accelerated epoch is not stable.

We can therefore draw on the $(r,m)$ plane the {}``forbidden direction
regions'' around the critical points, i.e. the direction for a curve
$m(r)$ intersecting the line $m=-r-1$ that must not be realized
(see Fig.~\ref{mr1} where we plot several possible $m(r)$ that
belongs to four general classes as detailed below). So, for any given
$m(r)$ model, one has simply to look at the intersections of $m(r)$
with $m=-r-1$ to decide if that model passes the conditions for a
standard matter-acceleration sequence. Generally speaking, if the
$m(r)$ line connects the standard matter era $(r,m)=(-1,0)$ with
an accelerated point $P_{6}$ or $P_{1}$ without entering the forbidden
direction region, then that model is cosmologically viable. Otherwise,
either because there is no connection at all or because the connection
has the wrong direction, then the model is to be rejected.

In general, of course, any $m(r)$ line is possible. However, assuming
$F>0$, one sees that $r(R)$ is a monotonic function, and therefore
$m(r)$ is single-valued and non-singular (remember we are assuming
a regular $f$ with all its derivatives). This simple property is
what we need to demonstrate our claims. In fact, it is then simple
to realize by an inspection of Fig.~\ref{mr1} that indeed it is
impossible to connect points near $m=+0$ with points in (A), (B)
or (C). To do so it would require in fact either entering the forbidden
direction regions or a turn-around of $m(r)$, i.e. a multi-valued
function, or a singularity of $m$ at finite $r$, or finally a crossing
of the critical line. This simple argument shows that the matter era
with $m\approx+0$ cannot connect to $P_{6}$ in the region (A), (B)
or (C). Hence the only accelerated point left is $P_{1}$ (which is
stable only for $0 \le m(r=-2)\le1$). Notice that this argument applies
for any number of roots in (A), (B) or (C).

A connection to $P_{6}$ is however possible at $r\to\pm\infty$,
with slope $m_{6}'=-1$, i.e. when the curve $m(r)$ is asymptotically
convergent on the $m=-r-1$ line. Even in this case, the final acceleration
is de-Sitter, although with $(x_{1},x_{2},x_{3})=(-1,0,2)$ instead
of $P_{1}:(x_{1},x_{2},x_{3})=(0,-1,2)$. To complete this demonstration
we need also to ensure that although the $m(r)$ line can have any
number of intersection with the critical line, no cosmological trajectory
can actually cross it. This property is indeed guaranteed by Eq.~(\ref{eq:critline}):
trajectories stop at the intersections of $m(r)$ with the critical
line and remain trapped between successive roots.

\subsubsection{From $P_{5}$ $(m<0)$ to $P_{6}$ $(m_{6}'>-1)$ in the region (B)}

There is then a further option: $P_{5}^{(0)}$ in the (B) region,
i.e. $m_{5}<0$. When $m$ is close to $-0$, one of the last two
eigenvalues in Eq.~(\ref{eig}) is positive whereas another is negative.
This shows that in this case the point $P_{5}^{(0)}$ is a saddle
independently of $m_{5}'$. Note that the accelerated point $P_{6}$
in the region (B) is stable for $m_{6}'>-1$.

Let us first consider the case $m_{5}'>-1$. Then the same argument
applies for the positive $m$ case discussed above. The $m(r)$ curves
can not satisfy both the conditions $m_{5}'>-1$ and $m_{6}'>-1$
required for the existence of the stable accelerated point $P_{6}$
in the regions (A), (B) and (C). However there is one exception. If
the matter root $m$ is small and strictly negative and $m_{5}'>-1$,
then $P_{6}$ for the \emph{same} root lies in the (B) region and
is a valid acceleration point. In other words, $P_{5}$ and $P_{6}$
coincide in the $(r,m)$ plane and are both acceptable since $m_{5,6}'>-1$.
The simplest possibility is $m={\rm const}\in(-1/2,0)$. For instance,
for the power-law models $f(R)=\alpha R^{0.9}$ (i.e. $m=-0.1$),
the transition from an approximately matter epoch $P_{5}^{(0)}$ to
an accelerated era $P_{6}$ is possible. However the matter period
is short because of real eigenvalues which diverge in the limit $m\to-0$,
Another possibility is $m=a+br$, i.e. a straight line intersecting
the critical line at some point with abscissa $(a-b)/(1+b)\in(-1/2,0)$
and a slope $b>-1$.

When $m_{5}'<-1$ it is possible to reach the stable accelerated point
$P_{6}$ in either of the regions (A), (B), (C) with $m_{6}'>-1$.
However the matter epoch does not last long in this case as well because
we have seen an eigenvalue is very large. Moreover, by construction
there will always be the final attractor $P_{6}$ in the region (B)
for the same $m$, whose effective equation of state corresponds to
a strongly phantom ($w_{{\rm eff}}<-7.6$).

Thus if the matter point $P_{5}^{(0)}$ exists in the region $m<0$,
the models are hardly compatible with observations because the matter
era is practically absent and because most trajectories will fall
in a unacceptable strongly phantom era.

\subsubsection{From $P_{5}$ $(m_{5}'>-1,m>0)$ to $P_{6}$ $(m_{6}'<-1)$ in the
region (D)}

\begin{figure}
\includegraphics[scale=1.2]{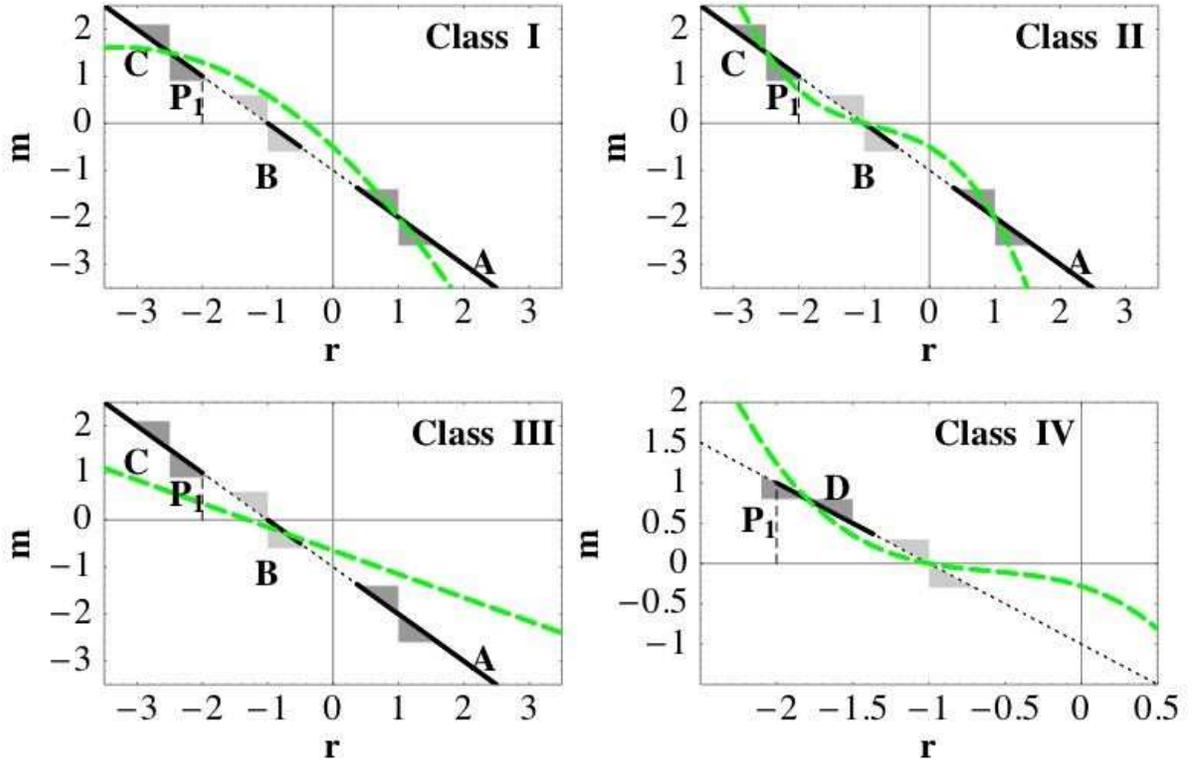}

\caption{\label{mr1}The $(r,m)$ plane for the four classes of $f(R)$ models.
In all panels, the straight diagonal line is the critical line $m=-r-1$.
In the dotted ranges $P_{6}$ is not accelerated or is unstable, if
we assume $m'_{6}>-1$. In the thick ranges labelled by A,B and C,
$P_{6}$ is accelerated and stable, again assuming $m'_{6}>-1$ (we
omit the region D for clarity except for the Class IV panel). The
gray triangles represent the forbidden directions near the critical
points. The dashed green lines are hypothetical $m(r)$ curves, intersecting
the critical line in the critical points $P_{5}$ and $P_{6}$. The
intersection at $(r,m)=(-1,0)$ (light gray triangles) corresponds
to the standard matter epoch $P_{5}^{(0)}$. In Class I models, the
$m(r)$ curve does not intersect $(r,m)=(-1,0)$ and therefore there
is no standard matter era. In Class II models, the point $(r,m)=(-1,0)$
is connected to the $P_{1}$ de-Sitter point (along the segment $0<m\le1$
at $r=-2$) and therefore represents a viable cosmological solution.
The two additional critical points in the regions A and C are unstable
since the curve enters the forbidden triangles and are therefore not
acceptable as final accelerated stages. In Class III models the $m(r)$
line with a slope $m'>-1$ intersects the critical line at a negative
$m$ in the strongly phantom range (B). Note that the curves with
$m_{5}'<-1$ which are attracted by $P_{6}$ in the region (A), (B),
(C) are possible, but such cases are not viable because of the absence
of a prolonged matter era for $m<0$. In Class IV models, the $m(r)$
curve connects the matter era with $m_{5}'>-1$ to the region (D)
with a derivative $m_{6}'<-1$ and therefore represents a viable cosmology
with a matter era followed by a stable acceleration ($w_{{\rm eff}}>-1$).
No single trajectory can cross the critical line $m=-r-1$: each solution
is trapped between two successive roots on the critical line. }
\end{figure}

We come to the fourth range, i.e. the region (D). Now the situation
is different for the point $P_{6}$, since $m_{6}'$ has to be \emph{less}
than $-1$ in order to be stable. Then it is possible to leave the
matter epoch $P_{5}^{(0)}$ (which satisfies $m_{5}'>-1,m>0$) and
to enter the accelerated epoch $P_{6}$ ($m_{6}'<-1$) as we illustrate
in Fig.~\ref{mr1} (Class IV panel). Therefore these models are compatible
with standard cosmology: they have a matter era followed by a non-phantom
acceleration with $w_{{\rm eff}}>-1$. Note that the saddle matter
epoch needs to be sufficiently long for structure formation to occur.
Later we shall provide an example of such models.

\vspace{0.5cm}
 Finally, we must mention an exception to this general argument. If
the $m(r)$ line has a derivative exactly $m'=-1$ at the critical
point, then that point is marginally stable and our linearized analysis
breaks down. In this case, one has to go to a second-order analysis
or to a numerical study. We will encounter such a situation for the
model $f(R)=R\log(\alpha R)^{q}$ we study later. The same applies
if $m'\to\pm\infty$, i.e. for trajectories that lie on the borders
of the forbidden regions.

\subsection{Classification of $f(R)$ models}

These discussions show that we can classify the $f(R)$ models into
four classes, as anticipated in Introduction. The classification can
be based entirely upon the geometrical properties of the $m(r)$ curve
and applies to all the cases in which an accelerated attractor exists
(see Fig. \ref{mr1}).

\begin{figure}
\includegraphics[scale=1.2]{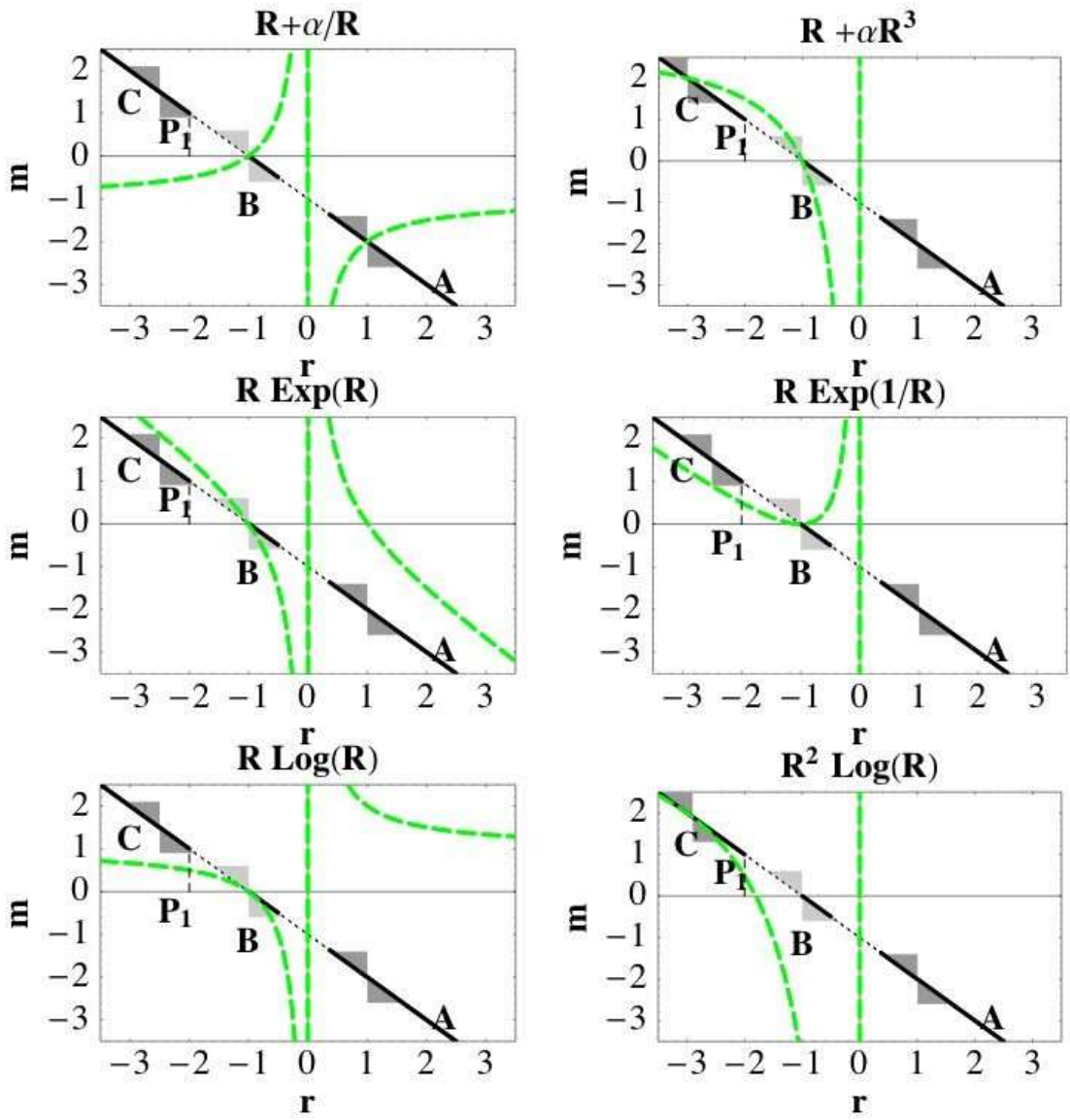}

\caption{\label{mr2}This figure illustrates several possible $m(r)$ curves
(thick dashed green line). Only the $f=R\log R$ and the $f=R\exp(1/R)$
models show an acceptable connection between the matter point $(r,m)=(-1,0)$
and the de-Sitter point $P_{1}$ along the dashed segment at $r=-2$.
In all other cases, there is either no intersection of the $m(r)$
curves with the critical line $m=-r-1$ near $(r,m)=(-1,0)$ or the
$m(r)$ curve enters the forbidden direction regions (the gray triangles).
In all panels we show the forbidden regions for three points in the
(A,B,C) ranges of $P_{6}$, even when there are no critical points
in one of those regions. For clarity we omit the range (D).}
\end{figure}

\begin{itemize}
\item Class I\,:\, This class of models covers all cases for which the
curve $m(r)$ does not connect the accelerated attractor with the
standard matter point $(r,m)=(-1,0)$, either because $m(r)$ does
not pass near the matter point, i.e. $m(r\to-1)\not=0$ , or because
the branch of $m(r)$ that accelerates is not connected to $(r,m)=(-1,0)$.
Instead of having a standard matter phase, the solutions reach the
$\phi$MDE fixed point $P_{2}$ with a {}``wrong'' evolution of
the scale factor ($a\propto t^{1/2}$) or bypass it altogether by
falling on the final attractor without a matter epoch at all. The
final accelerated fixed points, if they exist, can be in any of the
three ranges of $P_{6}$. 
\item Class II\,:\, For these models the $m(r)$ curve connects the upper
vicinity of the point $(r,m)=(-1,0)$ (with $m>0$ and $m_{5}'>-1$)
to the point $P_{1}$ located on the segment $0 \le m\le1$ at $r=-2$,
or asymptotically to $P_{6}(r\to\pm\infty)$. Since the approach to
$P_{5}$ is on the positive side of $m$, the trajectory exhibits
a damped oscillation around the matter point {[}see Eq.~(\ref{eig})],
which is followed by the de-Sitter point $P_{1}$ or $P_{6}(r\to\pm\infty$).
Models of the Class II are observationally acceptable and the final
acceleration corresponds to a de-Sitter expansion. 
\item Class III: For these models the $m(r)$ curve intersects the critical
line at $-1/2<m<0$ (i.e. region B). In all these cases the approximated
matter era is a very fast transient and only a narrow range of initial
conditions may allow it. Generically, the matter era is followed by
a strongly phantom acceleration, although one could design models
with the other ranges of the critical line. The closer to a standard
matter epoch, the more phantom the final acceleration is ($w_{{\rm eff}}\to-\infty$
as $m\to-0$). Since the matter era is practically unstable and the
highest effective equation of state is $-w_{{\rm eff}}=7.6$ (which
implies $w_{{\rm DE}}\simeq w_{{\rm eff},0}/\Omega_{{\rm DE},0}$
even smaller), these models are generally ruled out by observations
(although a more careful numerical analysis is required). 
\item Class IV\,:\, For these models the $m(r)$ curve connects the upper
vicinity of the point $(r,m)=(-1,0)$ (with $m'_{5}>-1$, $m>0$)
to the region (D) located on the critical line $m=-r-1$ (with $m_{6}'<-1$).
These models are observationally acceptable and the final acceleration
corresponds to a non-phantom effective equation of state ($w_{{\rm eff}}>-1$). 
\end{itemize}
In Fig.~\ref{mr2} we show a gallery of $m(r)$ curves for various
$f(R)$ models. The above discussions clarify the conditions for which
$f(R)$ dark energy models are acceptable. Only the Class II or Class
IV models are in principle cosmologically viable. However we need
to keep in mind that what we have discussed so far corresponds to
the behavior only around critical points. One cannot exclude the possibility
that single trajectories with some special initial conditions happen
to reproduce an acceptable cosmology. It is therefore necessary to
confirm our general analysis with a thorough numerical check; by its
nature, this check can only be done on a case-by-case basis, and to
this we turn our attention in the next sections.

\section{Specific models: analytical results}

In this section we shall consider a number of $f(R)$ models in which
$m$ can be explicitly written in terms of the function of $r$ and
study the possibility to realize the matter era followed by a late-time
acceleration. Most of the relevant properties of these models can
be understood by looking at the $m(r)$ curves of Fig.~\ref{mr2}.

\subsubsection{$f(R)=\alpha R^{-n}$}

This power-law model gives a constant $m$ from Eq.~(\ref{mdef}),
namely \begin{equation}
m=-n-1\,,\label{powerre}\end{equation}
 where $r=n$. The curve $m(r)$ degenerates therefore to a single
point and this case reduces to a two-dimensional system in the absence
of radiation because of the relation $x_{3}=nx_{2}$. Hence the condition
$m(r=-1)=0$ is satisfied only for $n=-1$, i.e. Einstein gravity.
Since the initial conditions around the end of the radiation era are
given for positive $R$, the positivity of the term $f_{,R}=-n\alpha R^{-n-1}$
requires that $\alpha<0$ for $n>0$ and $\alpha>0$ for $n<0$. From
Eq.~(\ref{powerre}) one has $m>0$ for $n<-1$ and $m<0$ for $n>-1$.
Since the eigenvalues of $P_{5}$ for this model are given by the
latter two in Eq.~(\ref{P5eig}) \cite{APT2}, the matter point $P_{5}$
is a stable spiral for $n<-1$ (around $m\to+0$). Then the solutions
do not leave the matter era for the late-time acceleration.

On the other hand, $P_{5}$ is a saddle point for $-1<n<-0.713$ while
the $\phi$MDE point $P_{2}$ is stable in the overlapping range $-1<n<-3/4$.
However one of the eigenvalues of $P_{5}$ exhibits a positive divergence
in the limit $m\to-0$, which means that the matter point becomes
repulsive if $m$ is very close to $-0$. As we anticipated, in the
region around $m=-0$ the effective equation of state for $P_{6}$
corresponds to the strongly phantom type ($w_{{\rm eff}}<-7.6$),
i.e., to our Class III models. \cite{APT2}. The above discussion
shows that the saddle point $P_{5}$ is connected to either the $\phi$MDE
point $P_{2}$ or the strongly phantom point $P_{6}$. The more one
tries to get a standard matter era for $n\to-1$, the more phantom
becomes the final acceleration and the more divergent becomes the
eigenvalues. Moreover if we take into account radiation, the solutions
tend to stay away from the point $P_{5}$.

So the models of this type are always in Class I except for (i) $-1<n<-0.713$
(Class III) and for (ii) $-1.327<n<-1$ (they are asymptotically not
accelerated). Similar conclusions were found in Ref. \cite{clift}.

The pure power-law models correspond to points $(r=-n,m=-1-n)$ in
the $(r,m)$ plane. We can notice that the $\Lambda$CDM model $f=R-\Lambda$
corresponds to the horizontal line $m=0$, which connects the matter
era at $(r,m)=(-1,0)$ with the de-Sitter acceleration $P_{1}$ at
$(r,m)=(-2,0)$ and is therefore a valid Class II model. A possible
generalization of $\Lambda$CDM is given by the models \begin{equation}
f(R)=(R^{b}-\Lambda)^{c}\,,\end{equation}
 which generate a tilted straight line $m(r)=r(1-c)/c+b-1$. If the
intersection $m=-1+bc$ with the critical line is at $0<m\ll1$ and
the slope is given by $-1<(1-c)/c<0$, then the matter era is connected
with $P_{1}$ and the model is acceptable (Class II).

\subsubsection{$f(R)=R+\alpha R^{-n}$}

This model was proposed in Refs.~\cite{Capo,Carroll} to give rise
to a late-time acceleration. From Eqs.~(\ref{mdef}) and (\ref{ldef})
we obtain \begin{equation}
m(r)=-\frac{n(1+r)}{r}\,.\label{mre1}\end{equation}
 Notice that $m(r)$ is independent of $\alpha$. Since $m(r=-1)=0$
the models satisfy the necessary condition for the existence of the
matter point $P_{5}$.

Let us analytically study the attractor behavior of the model in more
details. Substituting Eq.~(\ref{P5P6}) for Eq.~(\ref{mre1}), we
find the solutions $m_{a}=0$ or $m_{b}=-(n+1)$, which holds for
the points $P_{5}$ and $P_{6}$. In this case the points $P_{5}$
and $P_{6}$ are characterized by \begin{eqnarray}
 &  & P_{5a}:\quad\left(0,-\frac{1}{2},\frac{1}{2}\right),\quad\Omega_{\m}=1,\quad w_{{\rm eff}}=0\,,\\
 &  & P_{5b}:\quad\left(-\frac{3(n+1)}{n},\frac{4n+3}{2n^{2}},\frac{4n+3}{2n}\right)\,,\quad\Omega_{\m}=-\frac{8n^{2}+13n+3}{2n^{2}}\,,\quad w_{{\rm eff}}=-1-\frac{1}{n}\,,\\
 &  & P_{6b}:~~\left(-\frac{2(n+2)}{2n+1},\frac{4n+5}{(n+1)(2n+1)},\frac{n(4n+5)}{(n+1)(2n+1)}\right)\,,\quad\Omega_{\m}=0\,,\quad w_{{\rm eff}}=-\frac{6n^{2}+7n-1}{3(n+1)(2n+1)}\,.\label{P6po}\end{eqnarray}
 Note that for $m_{a}=0$, $P_{6a}$ goes to infinity. We are interested
in the case where a (quasi) matter era is realized around $m_{5}\approx0$.

This family of models splits into three cases: 1) $n<-1$, 2) $-1<n<0$,
and 3) $n>0$. The intermediate cases $n=0,1$ are of course trivial.

\begin{itemize}
\item Case 1 ($n<-1$).

Since $m'=n/r^{2}$ we see that $m'(-1)<-1$ and therefore the matter
epoch around $m\approx+0$ is stable and no acceleration is found
asymptotically ($P_{1}$ is stable as well for $-2<n<0$). The case
$n=-2$ corresponds to Starobinsky's inflation model and the accelerated
phase exists in the asymptotic past rather than in the future. This
case does not belong to one of our main classes since there is no
future acceleration.

\item Case 2 ($-1<n<0$).

Then the condition at $r=-1$ is fulfilled for $R\to\infty$, and
we see that $m=n(n+1)\alpha R^{-n-1}/(1-n\alpha R^{-n-1})$ approaches
zero from the positive side if $\alpha<0$. In this case, there are
damped oscillations around the standard matter era and the final stable
de-Sitter point $P_{1}$ can be reached ($P_{6}$ is unstable): this
is the Class II model. Notice that $F<0$ for small $R$, but $F>0$
along the cosmologically acceptable trajectory. When $\alpha>0$,
two of the eigenvalues diverge as $m\to-0$ and the matter era becomes
unstable.

\item Case 3 ($n>0$).

In this case the stable accelerated point $P_{6}$ exists in the non-phantom
region (A) because of the condition $m=-n-1<-1$. If $\alpha>0$,
$m$ approaches zero from the positive side. Then there are oscillations
around the matter era but the accelerated point $P_{1}$ is unstable
(since $m_{1}=-n/2<0$). Since $m_{5}'=n>0$, the matter era corresponds
to a saddle. However $P_{5}$ with $m_{5}'>-1$ cannot be connected
to $P_{6}$ in the region (A), as we showed in the previous section.
Hence we do not have a stable accelerated attractor after the matter
epoch. When $\alpha<0$, $m$ approaches zero on the negative side
and here again the matter point becomes effectively unstable since
one of the eigenvalues exhibits a positive divergence. Then this case
does not possess a prolonged matter epoch and belongs to the Class
I. The first panel of Fig.~\ref{mr2} shows graphically why models
like $f(R)=R+\alpha/R$ cannot work as a viable cosmological model:
the accelerated point is disconnected from the matter point.

\end{itemize}
In the next section we shall numerically confirm that the matter phase
is in fact absent prior to the accelerated expansion except for models
$f(R)=R+\alpha R^{-n}$ with $\alpha<0$ and $-1<n<0$. In any case,
all these power-law cases are cosmologically unacceptable. These results
fully confirm the conclusions of Ref.~\cite{APT} reached by studying
the Einstein frame. The single exception pointed out above for $-1<n<0$
was not a part of the cases considered in Ref.~\cite{APT}, since
$F<0$ for small $R$.

\subsubsection{$f(R)=R^{p}\exp(qR)$}

In this model $m$ is given by \begin{equation}
m(r)=-r+\frac{p}{r}\,.\label{m2re}\end{equation}
 Notice that for the pure exponential case ($p=0$) we have $m=-r$
and $x_{3}/m\to x_{2}\to0$ so that $P_{2}$ exists while $P_{5}$
does not. Otherwise the function $m$ vanishes for $r\to\pm\sqrt{p}$,
which means that the condition (\ref{m5con}) for the existence of
the matter era holds only for $p=1$. However, since in this case
$m'(r=-1)=-2<-1$ , the point $P_{5}$ is a stable spiral for $m>0$.
So the entire family of models is in fact ruled out.

In the limit $m\to+0$, $P_{6}$ can not be used for the late-time
acceleration in addition to the fact that $P_{5}$ is stable. Moreover
since $m(r=-2)=3/2$ for $p=1$, the de-Sitter point $P_{1}$ is not
stable. We note that Eqs.~(\ref{m2re}) and (\ref{P5P6}) are satisfied
in the limit $m\to+\infty$ and $r\to-\infty$, see Fig.~\ref{mr2}.
Since the eigenvalues in Eq.~(\ref{eigenP6}) are $-4,-4,0$ in this
case, the point $P_{6}$:\,$(x_{1},x_{2},x_{3})=(-1,0,2)$ with $m\to+\infty$
is marginally stable with an effective equation of state $w_{{\rm eff}}\to-1$.
In fact, when $m>0$, we have numerically checked that the final attractor
is either the matter point $P_{5}$ or $P_{6}:\,(x_{1},x_{2},x_{3})=(-1,0,2)$
(but then without a preceding matter phase), depending upon initial
conditions.

Thus models of this type do not have the sequence of matter and acceleration
for $p=1$, whereas the models with $p\neq1$ belong to Class I.

\subsubsection{$f(R)=R^{p}(\log\alpha R)^{q}$}

In this model we obtain the relation \begin{equation}
m(r)=\frac{p^{2}+2pr-r(q-r+qr)}{qr}\,.\label{mrlog}\end{equation}
 Since $m(r=-1)=-(p-1)^{2}/q$, the matter epoch exists only for $p=1$.
When $p=1$ one has $m(r=-2)=1-1/(2q)$, which means that $P_{1}$
is stable for $q>0$ whereas it is not for $q<0$. The derivative
term $m'(r)$ is given by \begin{equation}
m'(r)=-1+\frac{r^{2}-1}{qr^{2}}\,.\label{mrre}\end{equation}
 Since $m'(r=-1)=-1$ the point $P_{5}$ is marginally stable. However
we have to caution that $m$ does not exactly become zero. In fact
when $r<-1$ we have $m'(r)>-1$ and $m(r)>0$ for $q>0$, which means
that the quasi matter era with positive $m$ is a saddle point. Similarly
the accelerated point $P_{6}$ in the region (C) is stable for $q>0$
whereas it is not for $q<0$. Hence both $P_{1}$ and $P_{6}$ are
stable for positive $q$. However one can show that the function $m(r)$
given in Eq.~(\ref{mrlog}) satisfies $m(r)<-r-1$ in the region
$r<-1$ for $p=1$ and $q>0$. Hence the curve (\ref{mrlog}) does
not cross the point $P_{6}$ in the region (C). Then the only possibility
is the case in which the trajectories move from the quasi matter era
$P_{5}$ to the de-Sitter point $P_{1}$. In the next section we shall
numerically show that the sequence from $P_{5}$ to $P_{1}$ is in
fact realized.

Thus when $p=1$ and $q>0$ the above model corresponds to the Class
II, whereas the models with $p\neq1$ are categorized as the Class
I.

\subsubsection{$f(R)=R^{p}\exp(q/R)$}

This model gives the relation \begin{equation}
m(r)=-\frac{p+r(2+r)}{r}\,,\label{mrexp}\end{equation}
 which is independent of $q$. Here we have $m(r=-1)=p-1$, so a matter
era exists for $p=1$. In this case one has $m(r)=-(r+1)^{2}/r>0$
for $r<0$. Since $m(r=-2)=1/2$ for $p=1$, the point $P_{1}$ is
a stable spiral. The derivative term $m'(r)$ is given by $m'(r)=-1+1/r^{2}$,
which then implies $m'(r=-1)=0$ and $m'(r<-1)>-1$. This shows that
$P_{5}$ is a saddle whereas $P_{6}$ in the region (C) is stable.
The curve (\ref{mrexp}) satisfies the relation $m(r)<-r-1$ in the
region $r<-1$ for $p=1$ and also has an asymptotic behavior $m(r)\to-r$
in the limit $r\to-\infty$. Then in principle it is possible to have
the sequence $P_{5}\to P_{6}(r\to-\infty)$, but the trajectory from
the point $P_{5}$ is trapped by the stable de-Sitter point $P_{1}$
which exists at $(r,m)=(-2,1/2)$. We note that one of the eigenvalues
for the point $P_{5}$ is large ($3(1+m_{5}')=3$) compared to the
model $f(R)=R(\log\alpha R)^{q}$ whose eigenvalue is close to 0 (but
positive) around $m=0$. In such a case the system does not stay around
the matter point $P_{5}$ for a long time as we will see later.

Thus the model with $p=1$ belongs to the Class II, whereas the models
with $p\neq1$ correspond to the Class I.

\subsubsection{$f(R)=R+\alpha R^{2}-\Lambda$}

In this case the function $m(r)$ is given by \begin{equation}
m(r)=\frac{-1-r+A(r)}{1+A(r)}\,,\label{eq:case6}\end{equation}
 where \begin{equation}
A(r)\equiv\sqrt{(1+r)^{2}+4\tilde{\alpha}r(2+r)}\,,\quad\tilde{\alpha}\equiv\alpha\Lambda\,.\end{equation}
 Here we assume that $\alpha,\Lambda>0$. The equation, $m(r)=-1-r$,
gives three solutions \begin{eqnarray}
r_{1,2}=-\frac{1+4\tilde{\alpha}\pm2B}{1+4\tilde{\alpha}}\,,\quad r_{3}=-2\,,\end{eqnarray}
 where $B\equiv\sqrt{\tilde{\alpha}(1+4\tilde{\alpha})}$. Then we
obtain three points $P_{5}$ and three $P_{6}$. For $P_{5}$ we have
\begin{eqnarray}
 &  & P_{5a,b}:(x_{1},x_{2},x_{3})=\left(\frac{6\tilde{\alpha}}{2\tilde{\alpha}\pm B},-\frac{B(B\pm8\tilde{\alpha})}{2(B\pm2\tilde{\alpha})^{2}}\,,\frac{8\tilde{\alpha}\pm B}{4\tilde{\alpha}\pm2B}\right)\,,\\
 &  & P_{5,c}:(x_{1},x_{2},x_{3})=\left(\frac{3}{2},-\frac{5}{8},\frac{5}{4}\right)\,.\end{eqnarray}
 The point $P_{5,c}$ is unphysical since $\Omega_{\m}<0$. The points
$P_{5a,b}$ reduce to a matter point in the limit $\tilde{\alpha}\ll1$.
At the lowest order one has $w_{{\rm eff}}\approx\mp4\sqrt{\tilde{\alpha}}/3$.
This shows that a standard matter era can exist either for $\alpha\to0$,
i.e., for the $\Lambda$CDM model, or for $\Lambda\to0$, i.e., for
the Starobinsky's model $f(R)=R+\alpha R^{2}$. In the limit $\tilde{\alpha}\to0$
the only accelerated point is the de-Sitter point $P_{1}$. Since
the condition $m(-2)=1$ is satisfied for any $\tilde{\alpha}$, we
see that this $f(R)$ model is always attracted by the de-Sitter acceleration.

Models of this type belong to Class II.

\subsubsection{$f(R)=R-\mu_{1}^{4}/R+\mu_{2}^{-2}R^{2}$}

This model was proposed in Ref.~\cite{NO03}. In this case Eq.~(\ref{P5P6})
reads: \begin{equation}
R^{3}~\frac{2+r}{\mu_{2}^{2}}+R^{2}\,(1+r)+\mu_{1}^{4}\,(1-r)=0\,,\label{eq:case7}\end{equation}
 where $R$ needs to be real solutions. Since the solutions for this
equation are quite complicated, we will not write them down here.
The necessary condition for the existence of the matter phase is as
usual $m(-1)=0$. We have here \begin{equation}
m(-1)=\frac{6}{3-(2\mu_{2}/\mu_{1})^{4/3}}\,.\label{cond}\end{equation}
 Hence we see that $m(-1)$ tends to zero for $\mu_{1}\to0$ but since
it stays on the negative side, the matter era is unstable (one of
the eigenvalues exhibits a positive divergence). So we can draw from
Eq.~(\ref{cond}) an important conclusion that the matter phase can
only be obtained for $\mu_{1}=0$, i.e., the Starobinsky's (inflation)
model previously discussed.

In order to satisfy solar system constraints a particular version
of this model was suggested with \cite{NO03} \begin{equation}
\mu_{2}=3^{3/4}\mu_{1}\,,\quad{\rm and}\quad R=\sqrt{3}\mu_{1}^{2}\,.\end{equation}
 In that case, (\ref{cond}) yields $m(-1)\approx3.40$ hence this
case does not have a standard matter phase either.

The model has two accelerated attractors: \begin{equation}
\left\{ \begin{array}{ll}
P_{6}:(x_{1},x_{2},x_{3})=(-2,3/2,3/2)\,,\quad w_{{\rm eff}}=-2/3\,,\\
P_{1}:(x_{1},x_{2},x_{3})=(0,-1,2)\,,\quad w_{{\rm eff}}=-1\,.\end{array}\right.\end{equation}
 Thus depending upon the initial conditions the trajectories lie in
the basin of attraction of either of these two points.

This model corresponds to Class I.

\subsubsection{$m(r)=-0.2(1+r)(3.2+0.8r+r^{2})$}

This model has been designed by hand to meet the condition for the
Class IV. Note that this corresponds to the $m(r)$ curve in the Class
IV case shown in Fig.~\ref{mr1}. The corresponding $f(R)$ Lagrangians
are the solutions of the differential equation \begin{equation}
\frac{Rf_{,RR}}{f_{,R}}=m\left(-\frac{Rf_{,R}}{f}\right)\,,\end{equation}
 which can be obtained numerically. This model obeys the conditions
$m(-1)=0$ and $m_{5}'>-1$ required for a saddle matter era $P_{5}$
as well as the conditions $(\sqrt{3}-1)/2<m_{6}=0.8<1$ and $m_{6}'<-1$
required for a stable accelerated point $P_{6}$ in the region (D).
The final accelerated attractor corresponds to the effective equation
of state $w_{{\rm eff}}\approx-0.935$.

A model with similar properties but an analytical Lagrangian is $f(R)=R^{\frac{p}{p-1}}(R+C)^{\frac{1}{1-p}}$
($C\not=0,\, p\not=1$) for which $m(r)=-p(r+1)^{2}/r$, whose $P_{6}$
intersection lies in the region (D) for $2<p<3.73$ with $w_{{\rm eff}}=\frac{1-9p+2p^{2}}{3(1+p)}$.
Here however the matter era has large eigenvalues so in fact is of
very little duration and hardly realistic. A generalization to $m(r)=-p(r+r_{0})^{2}/r$
with $r_{0}$ slightly less than 1 works much better but then the
Lagrangian is very complicated.

\subsubsection{Summary}

In Table I we summarize the classification of most $f(R)$ dark energy
models presented in this section. No model belongs to the Class IV
except for the purposely designed cases given in the previous subsection,
so we omit the Class IV column. The models which are classified in
Class II at least satisfy the conditions to have a saddle matter era
followed by a de-Sitter attractor. This includes models of the type
$f=R+\alpha R^{-n}$ ($-1<n<0$, $\alpha<0$), $f=R(\log\alpha R)^{q}$
$(q>0)$ and $f=R+\alpha R^{2}-\Lambda$. However this does not necessarily
mean that these models are cosmologically viable, since it can happen
that the matter era is too short or too long to be compatible with
observations. In the next section we shall numerically study the cosmological
viability of the above models.

\begin{table}
\begin{tabular}{|c|c|c|c|c|}
\hline 
$f(R)$ models&
$m(r)$&
Class I&
Class II&
Class III\tabularnewline
\hline
\hline 
$\alpha R^{-n}$&
$-1-n$&
$n>-0.713$&
--&
$-1<n<-0.713$\tabularnewline
\hline 
$R+\alpha R^{-n}$&
$-n\frac{(1+r)}{r}$&
$n>0$&
$-1<n<0,\quad\alpha<0$&
--\tabularnewline
\hline 
$R^{p}(\log\alpha R)^{q}$&
$\frac{p^{2}+2pr-r(q-r+qr)}{qr}$&
$p\not=1$&
$p=1,\quad q>0$&
--\tabularnewline
\hline 
$R^{p}\exp qR$&
$-r+\frac{p}{r}$&
$p\not=1$&
--&
--\tabularnewline
\hline 
$R^{p}\exp(q/R)$&
$-\frac{p+r(2+r)}{r}$&
$p\not=1$&
$p=1$&
--\tabularnewline
\hline 
$R+\alpha R^{2}-\Lambda$&
Eq. (\ref{eq:case6})&
--&
$\alpha\Lambda\ll1$&
--\tabularnewline
\hline 
$R-\mu_{1}^{2}/R+R^{2}/\mu_{2}^{2}$&
Eq. (\ref{eq:case7})&
always&
--&
--\tabularnewline
\hline
\end{tabular}

\caption{Classification of $f(R)$ dark energy models.}
\end{table}

\section{Specific cases: Numerical results}

We will now use the equations derived in Section II in order to recover
the cosmic history of given $f(R)$ DE models and confirm and extend
our analytical results. In all cases, we include radiation and give
initial conditions at an epoch deep into the radiation epoch. As our
aim is to check their cosmological viability, we tune the initial
conditions in order to produce observationally acceptable values,
namely \begin{equation}
\Omega_{{\rm m,0}}\approx0.3\,,\qquad\Omega_{{\rm rad,0}}\approx10^{-4}\,.\label{par0}\end{equation}
 In some cases we plot a 2-dimensional projection of the 3-dimensional
phase space $(x_{1},x_{2},x_{3}$) (no radiation) in Poincaré coordinates,
obtained by the transformation $x_{i}^{({\rm P})}=x_{i}/(1+d)$ where
$d=\sqrt{x_{1}^{2}+x_{2}^{2}+x_{3}^{3}}$.

\subsection{$f(R)=\alpha R^{-n}$}

\begin{figure}
\includegraphics[width=3in,height=3in]{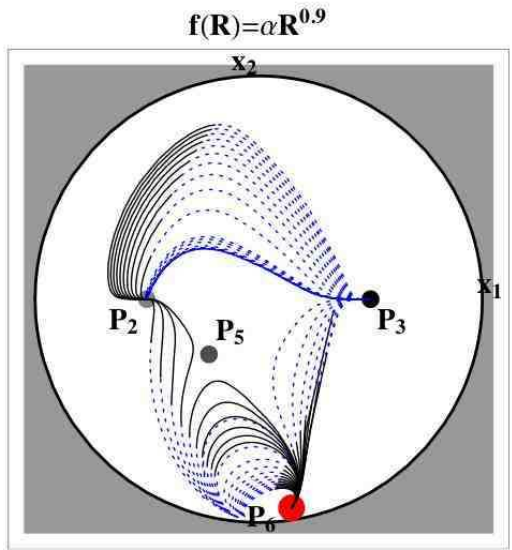}

\caption{\label{plot09} Phase space in the plane $(x_{1},x_{2})$ in Poincaré
coordinates for the model $f(R)=\alpha R^{0.9}$ in the absence of
radiation. Here and in the following plot, the dotted lines correspond
to trajectories at the early stage, the continuous lines to those
at the final stage. The circles represent critical points. The solutions
approach either the $\phi$MDE point $P_{2}$ or the phantom point
$P_{6}$. The point $P_{5}$ is a saddle, but the trajectories do
not approach this point if we take into account radiation. The point
$P_{3}$ is an unstable node. }
\end{figure}

Since $m=-n-1$ in this case, the matter era is possible only when
$n$ is close to $-1$. So let us consider the cosmological evolution
around $n=-1$. As we already showed, the matter point $P_{5}$ is
stable for $n<-1$. When $n>-1$, $P_{5}$ is a saddle and both $P_{2}$
and $P_{6}$ are stable. In Fig.~\ref{plot09} we show a 2-dimensional
phase space plot for the model $n=-0.9$ in the \textit{absence} of
radiation. In fact the final attractors are either the $\phi$MDE
point $P_{2}$ with $w_{{\rm eff}}=1/3$ or the phantom point $P_{6}$
with $w_{{\rm eff}}=-10.17$. The point $P_{5}$ with $w_{{\rm eff}}=1/9$
is in fact a saddle point. However, if we start from realistic initial
conditions around $(x_{1},x_{2},x_{3},x_{4})=(0,0,0,1)$ with the
inclusion of radiation, we have numerically found that the trajectories
directly approach final attractors ($P_{2}$ or $P_{6}$) without
reaching the vicinity of $P_{5}$. Moreover as we choose the values
of $n$ closer to $-1$, the point $P_{5}$ becomes repulsive because
of the positive divergence of an eigenvalue. These results show that
the power-law models with $n>-1$ do not provide a prolonged matter
era sandwiched by radiation and accelerated epochs in spite of the
fact that the point $P_{5}$ can be a saddle.

\subsection{$f(R)=R+\alpha R^{-n}$}

\begin{figure}
\includegraphics[width=3in,height=3in]{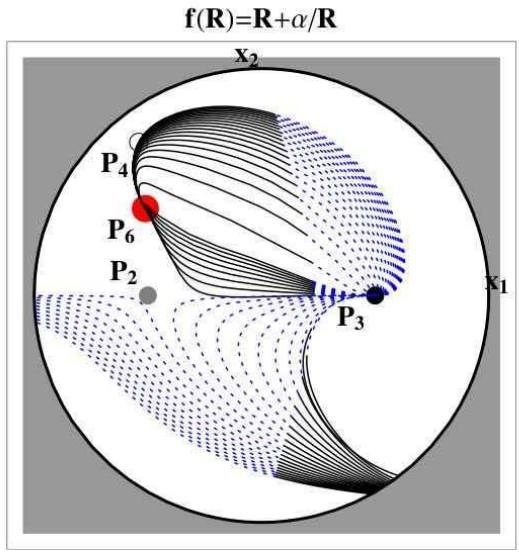}

\caption{\label{psn1} Phase space projected on the plane $(x_{1},x_{2})$
in Poincaré coordinates for the model $f(R)=R+\alpha/R$ in the absence
of radiation. For the initial conditions $x_{2}>0$ there are two
solutions: either (i) the solutions directly approach the accelerated
attractor $P_{6}$ or (ii) they first approach the saddle $\phi$MDE
point $P_{2}$ and then reach the attractor $P_{6}$. When $x_{2}<0$
initially, the trajectories moves toward $x_{2}\to-\infty$. Note
that the point $P_{3}$ is unstable.}
\end{figure}

When $n>0$ one has $m=-n-1<-1$ and $m'(r)=n/r^{2}>0$ for $P_{5}$
and $P_{6}$. In this case $P_{6}$ is a stable attractor whereas
$P_{5}$ is a saddle. In the previous section we showed that the matter
point $P_{5}$ is disconnected to the accelerated point $P_{6}$ since
$P_{6}$ exists in the region (A). According to the results in Ref.~\cite{APT}
we have only the following two cases: either (i) the matter era is
replaced by the $\phi$MDE fixed point $P_{2}$ which is followed
by the accelerated attractor $P_{6}$, or (ii) a rapid transition
from the radiation era to the accelerated attractor $P_{6}$ without
the $\phi$MDE. Which trajectories are chosen depend upon the model
parameters and initial conditions. In Fig.~\ref{psn1} we depict
a 2-dimensional phase space plot for the model $n=1$. This shows
that the final attractor is in fact $P_{6}$ and that whether the
solutions temporally approach the saddle point $P_{2}$ or not depends
on initial conditions.

\begin{figure}[H]

\begin{centering}\includegraphics{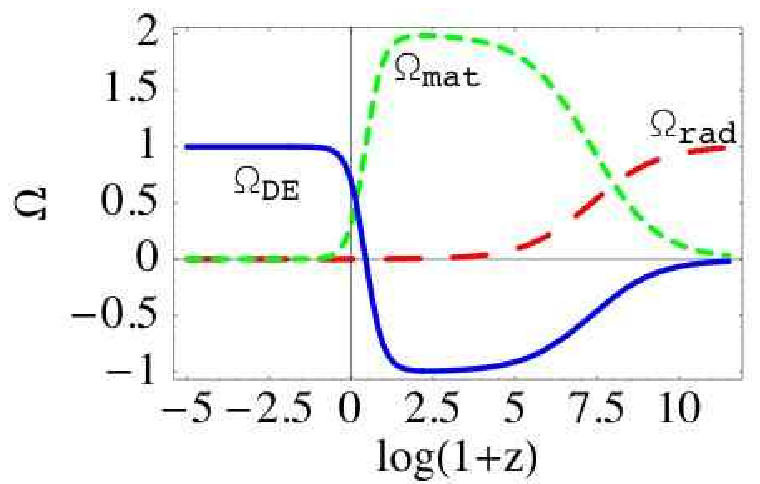} \includegraphics{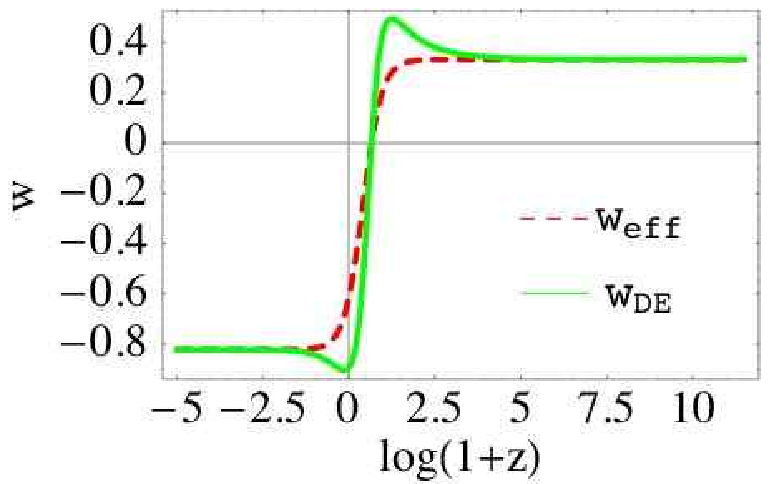} \par\end{centering}

\caption{The cosmic evolution of various quantities is shown for the model
$f(R)=R+\alpha/R^{2}$ with $\alpha=-\mu^{6}$, $\mu/H_{0}=11.04$.
The standard matter era is replaced by the $\phi$MDE which corresponds
to $a\propto t^{1/2}$, $w_{{\rm eff}}=1/3$ and $\Omega_{\m}=2$.
The redshift $z_{a}$ at which acceleration starts is $z_{a}=0.4$
and we have asymptotically in the future $\Omega_{{\rm DE}}=1$ and
$w_{{\rm eff}}=w_{{\rm DE}}=-0.82$ {[}see Eq.~(\ref{P6po})]. \label{evon2} }
\end{figure}

In order to understand the evolution after the radiation era let us
consider the model $n=1$ without radiation. From Eqs.~(\ref{E1})
and (\ref{E2}) we find that the evolution of the scale factor during
the $\phi$MDE is given by \begin{equation}
a(t)=(t/t_{i})^{1/2}+\epsilon(t)(t/t_{i})^{9/4}\,,\label{scalephi}\end{equation}
 where the subscript `\textit{i}' represents the value at the beginning
of the $\phi$MDE. At first order in $\epsilon(t)$ we have \begin{equation}
\epsilon(t)=\frac{\mu^{2}}{144H_{i}^{2}}\frac{1}{\sqrt{\rho_{m}^{(i)}/3H_{i}^{2}-(H/H_{i})^{1/2}}}\,.\end{equation}
 Notice that $\mu$ is of order $H_{0}$ to realize the present acceleration.
Since $H_{0}\ll H_{i}$ the parameter $\epsilon(t)$ is in fact much
smaller than unity. The scale factor evolves as $a\propto t^{1/2}$
during the $\phi$MDE, but this epoch ends when the second term in
Eq.~(\ref{scalephi}) gets larger than the zero-th order term. Hence
the end of the $\phi$MDE is characterized by \begin{equation}
t\approx\left(\frac{144H_{i}^{2}}{\mu^{2}}\sqrt{\frac{\rho_{m}^{(i)}}{3H_{i}^{2}}}\right)^{4/7}\, t_{i}\,.\label{tcrit}\end{equation}
 After that the solutions approach the accelerated attractor $P_{5}$.
Equation (\ref{tcrit}) shows that the duration of the $\phi$MDE
depends on $\mu$ together with the initial conditions $\rho_{m}^{(i)}$
and $H_{i}$. The similar argument can be applied for any $n<-1,n>-3/4$
with a correction growing as $t^{5/2-1/2(n+1)}$. In Fig.~\ref{evon2}
we plot the evolution of various quantities for $n=2$. In this case
the radiation era is followed by the $\phi$MDE saddle point $P_{2}$
with $\Omega_{\m}=2$ and $w_{{\rm eff}}=1/3$. The final attractor
is the accelerated point $P_{6}$ with $\Omega_{{\rm DE}}=1$ and
$w_{{\rm eff}}=-0.82$. As is clearly seen in the right panel of Fig.~\ref{evon2}
we do not have a standard matter era with $w_{{\rm eff}}=0$.

%
\begin{figure}
\includegraphics[width=3in,height=3in]{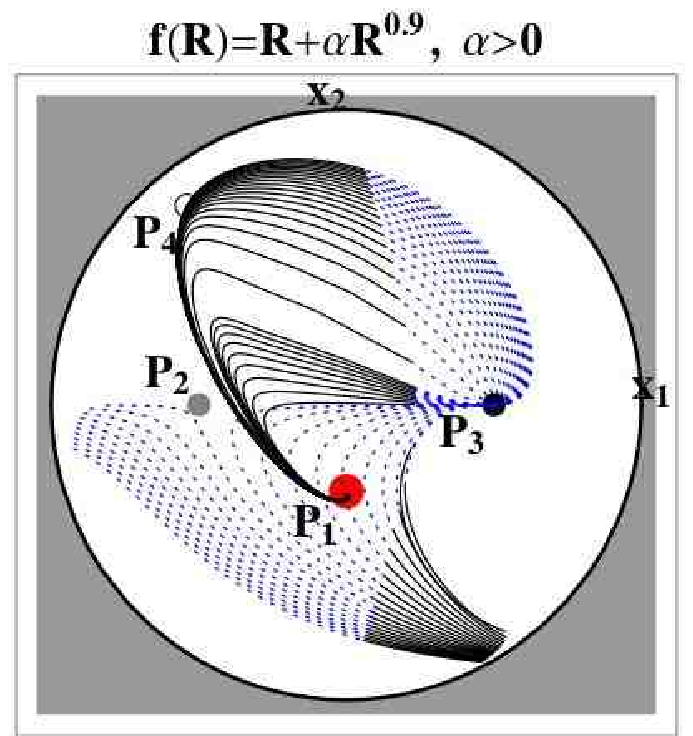} \includegraphics[width=3in,height=3in]{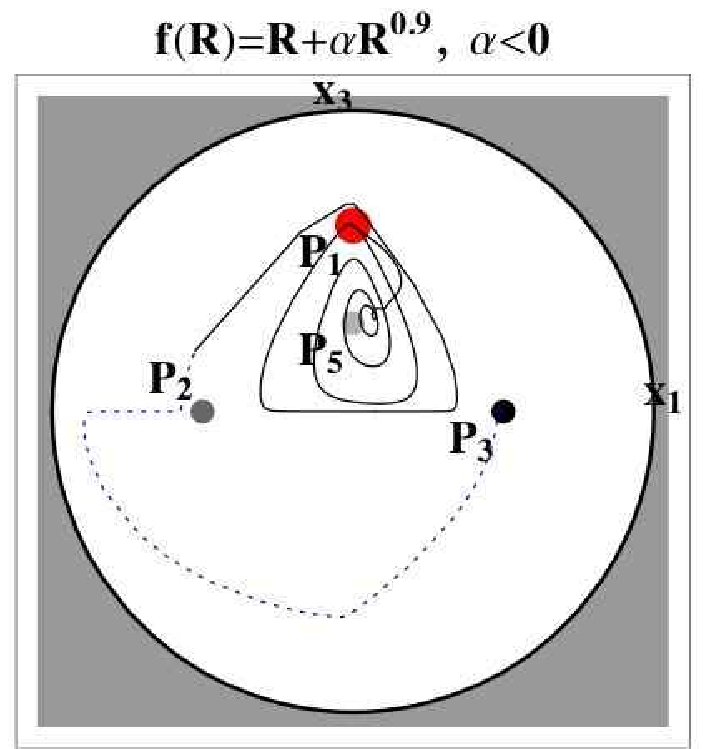}

\caption{\label{psn09} Projected phase space in Poincaré coordinates for
the model $f(R)=R+\alpha R^{0.9}$ in the absence of radiation. The
final attractor is the de-Sitter point $P_{1}:(x_{1},x_{2},x_{3})=(0,-1,2)$.
Note that both $P_{2}$ and $P_{6}$ are not stable unlike the model
$f(R)=\alpha R^{0.9}$. In the left panel, $\alpha>0$: here $P_{5}$
corresponds to $m<0$ with a large eigenvalue and therefore is unstable.
In the right panel, $\alpha<0$: now the point $P_{5}$ is a saddle
with positive $m$, so it is possible to have a sequence of an oscillating
matter phase followed by the late-time acceleration. We plot a single
curve for clarity.}
\end{figure}

Let us consider the case in which $n$ is close to $-1$. When $n<-1$
the point $P_{5}$ is a stable spiral, so the matter era is not followed
by an accelerated expansion as is similar to the power-law models.
If $n>-1$, the de-Sitter point $P_{1}$ is stable whereas the phantom
point $P_{6}$ is not. In Fig.~\ref{psn09} we show the phase space
plot in a two-dimensional plane for $n=-0.9$. When $\alpha>0$, although
the point $P_{5}$ is a saddle, the solutions approach the attractor
$P_{1}$ without staying the region around the point $P_{5}$ for
a long time because $m$ is negative. This tendency is more significant
if $n$ is chosen to be closer to $-1$, i.e. $m\to-0$. Hence one
can not have a prolonged matter era in these cases as well. On the
other hand, for $\alpha<0$, we have $m\to+0$ and there are oscillations
around the matter era followed again by the attractor $P_{1}$. Then
this latter case, belonging to Class II, can be cosmologically viable.

\subsection{$f(R)=R(\log\alpha R)^{q}$}

When $q>0$, we showed that the point $P_{5}$ is a saddle for $m(r<-1)>0$
whereas both $P_{1}$ and $P_{6}$ are stable. {}In the previous
section we showed that the only possibility is the trajectory from
$P_{5}$ to $P_{1}$. Hence the solutions starting from the radiation
era reach the saddle matter point $P_{5}$ first, which is followed
by the de-Sitter point $P_{1}$.

\begin{figure}[H]
\begin{centering}\includegraphics{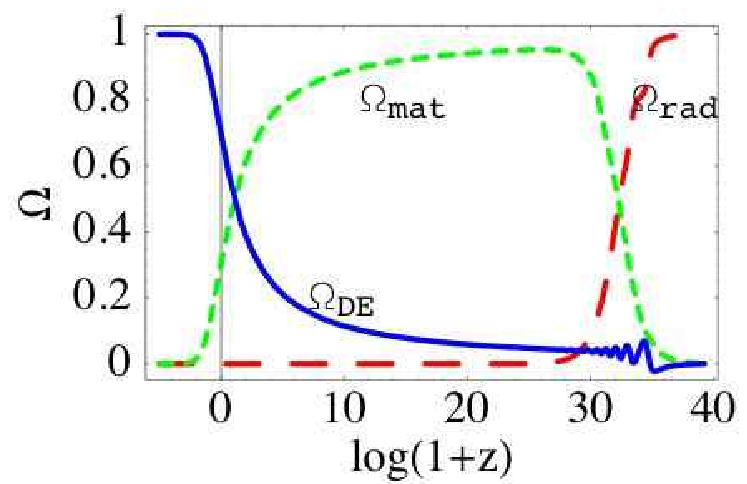} \includegraphics{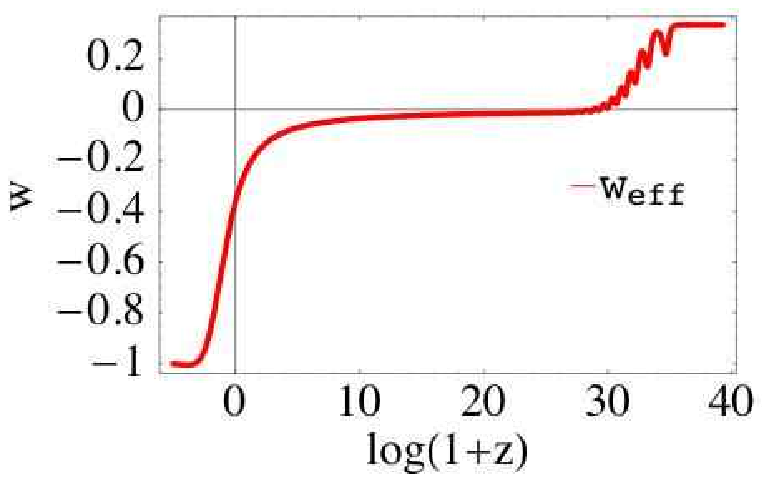} \par\end{centering}

\caption{The cosmic evolution of various quantities for the model $f(R)=R{\rm log}\,\alpha R$
with initial conditions $x_{1}=10^{-5}$, $x_{2}=-10^{-10}$, $x_{3}=1.01\times10^{-10}$
and $x_{4}=0.999$ at the redshift $z=1.1\times10^{17}$, corresponding
to $r=-1.01$. In this case the matter era is too long relative to
the standard cosmology. In fact the energy fraction of the radiation
at the present epoch is $\Omega_{{\rm rad},0}=2.8\times10^{-15}$,
which is much smaller than the standard value $\Omega_{{\rm rad},0}\approx10^{-4}$. }

\label{logmodel} 
\end{figure}

In order to obtain a prolonged matter period, the variables $m$ and
$r$ need to be close to $+0$ and $r=-1$, respectively, at the end
of radiation era. If we integrate the autonomous equations with initial
conditions $r=x_{3}/x_{2}\simeq-1$ (and smaller than $-1)$ and $x_{4}\simeq1$,
we find that the matter era is too long to be compatible with observations.
In Fig.~\ref{logmodel} we plot one example of such cosmological
evolution for $q=1$. This shows that a prolonged (quasi) matter era
certainly exists prior to the late-time acceleration. The final attractor
is the de-Sitter point $P_{1}$ with $w_{{\rm eff}}=-1$. However
in this case the beginning of the matter epoch corresponds to the
redshift $z=1.1\times10^{17}$, which is much larger compared to the
standard value $z\sim10^{3}$. The present value of the radiation
energy fraction is $\Omega_{{\rm rad},0}=2.8\times10^{-15}$ and is
much smaller than the value given in Eq.~(\ref{par0}).

\begin{figure}[H]
\begin{centering}\includegraphics{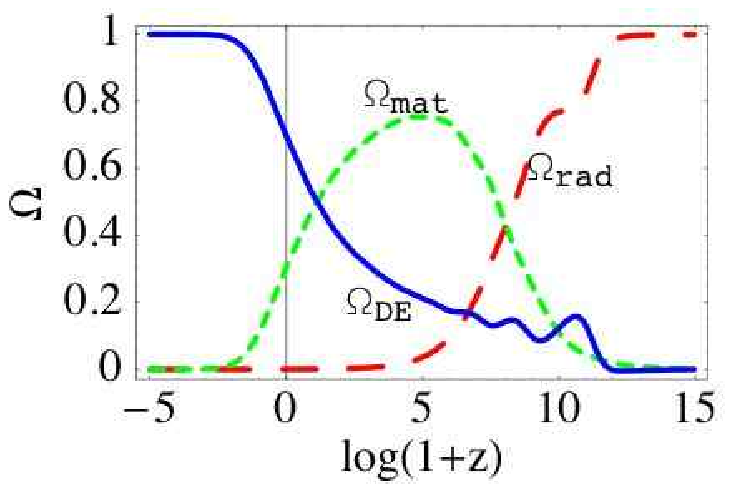} \includegraphics{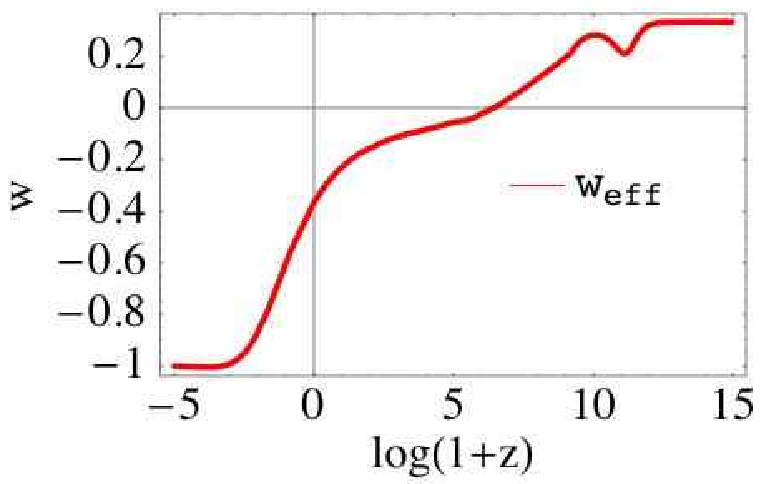} \par\end{centering}

\caption{The cosmic evolution of various quantities for the model $f(R)=R{\rm log}\,\alpha R$
with initial conditions $x_{1}=10^{-10}$, $x_{2}=-10^{-7}$, $x_{3}=1.019\times10^{-7}$
and $x_{4}=0.999$ at the redshift $z=3.15\times10^{6}$, corresponding
to $r=-1.019$. In this case we have $\Omega_{{\rm m,0}}\approx0.3$
and $\Omega_{{\rm rad,0}}\approx10^{-4}$ at the present epoch, but
the matter era is practically absent. }

\label{logmodel2} 
\end{figure}

This unusually long period of the matter era is associated with the
fact that the point $P_{5}$ is a saddle in the region $r<-1$ but
it is marginally stable in the limit $r\to-1$ (i.e. $m\to+0$). Hence
as we choose the initial values of $r$ closer to $-1$, the duration
of the matter period gets longer. In order to recover the present
value of $\Omega_{{\rm rad}}$ given in Eq.~(\ref{par0}), we have
to make the matter period shorter by appropriately choosing initial
conditions at the end of the radiation era. In Fig.~\ref{logmodel2}
we plot the cosmological evolution in the case where the end of the
radiation era corresponds to $z\sim10^{3}$ with present values $\Omega_{{\rm m,0}}\approx0.3$
and $\Omega_{{\rm rad,0}}\approx10^{-4}$. The energy fraction of
the matter is not large enough to dominate the universe after the
radiation epoch. Hence this case is not compatible with observations
as well.

\subsection{$f(R)=R\exp(q/R)$}

\begin{figure}[H]

\begin{centering}\includegraphics{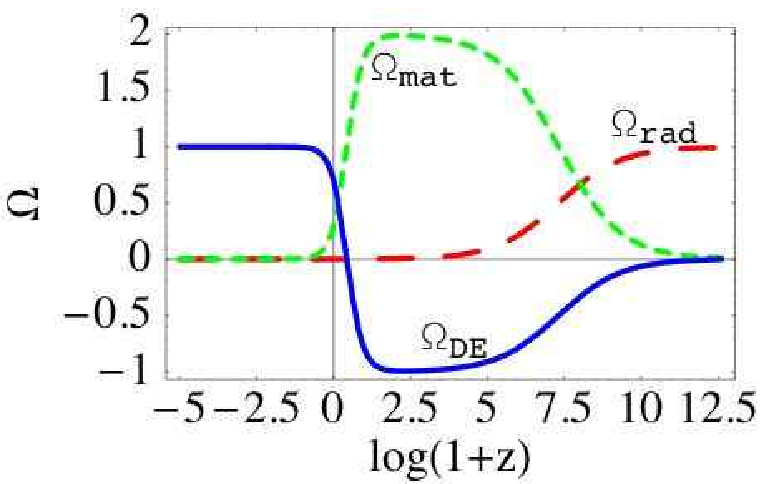} \includegraphics{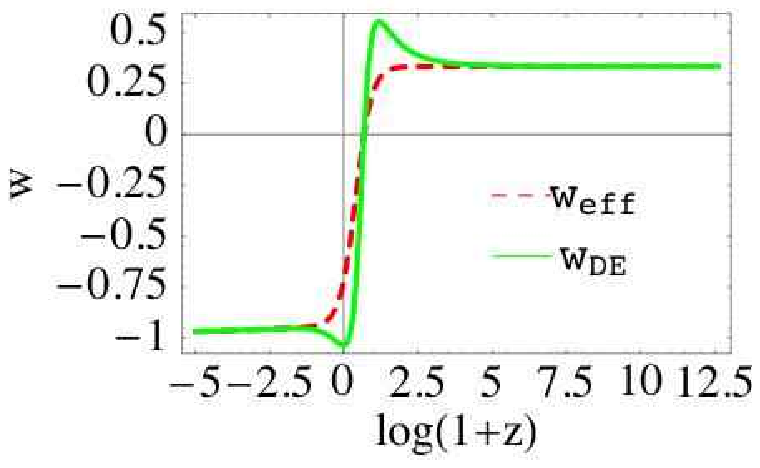} \par\end{centering}

\caption{The cosmic evolution of various quantities for the model $f(R)=R\exp(q/R)$
with initial conditions $x_{1}=0$, $x_{2}=2.13\times10^{-20}$, $x_{3}=5.33\times10^{-21}$
and $x_{4}=0.99$ at the redshift $z=3\times10^{5}$\textbf{.} We
see that the matter era is absent and is replaced by the $\phi$MDE.}

\label{Rexp} 
\end{figure}

In this case the matter point $P_{5}$ is a saddle, but one of the
eigenvalues are 3 rather than close to 0. Numerically we find that
the solutions do not reach the matter dominated epoch unlike the $f(R)=R(\log\alpha R)^{q}$
model with $q>0$. In Fig.~\ref{Rexp} we plot the cosmological evolution
for this model corresponding to the present values $\Omega_{{\rm m,0}}\approx0.3,\Omega_{{\rm rad,0}}\approx10^{-4}$.
In this case the matter epoch is replaced by the $\phi$MDE. It is
possible to find a situation in which there exists a short period
of the matter era, but we find that this case does not satisfy the
conditions given by (\ref{par0}). Thus this model is not cosmologically
viable in spite of the fact that it belongs to the Class II.

\subsection{$m(r)=-0.2(1+r)(3.2+0.8r+r^{2})$}

\begin{figure}[H]
\begin{centering}\includegraphics{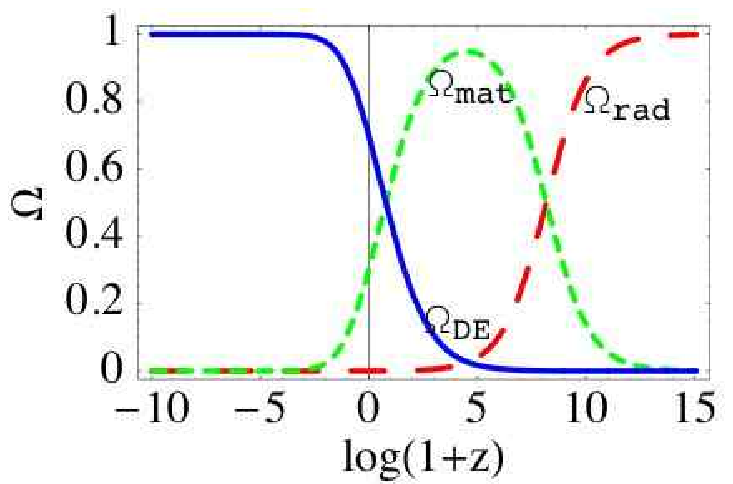} \includegraphics{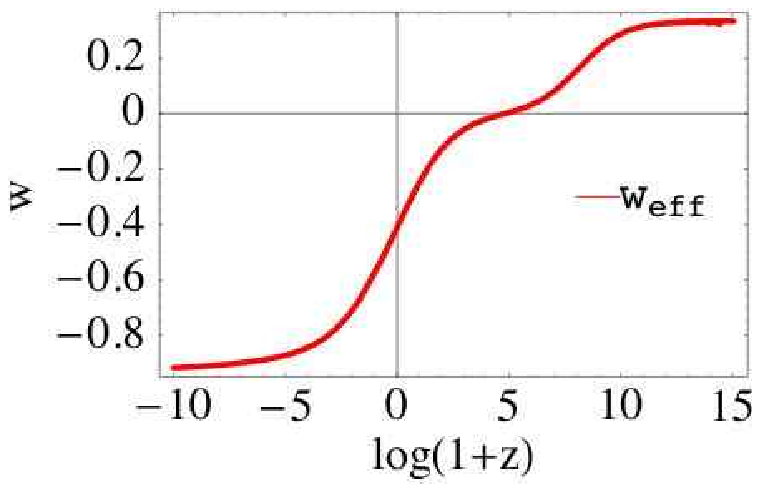} \par\end{centering}

\caption{The cosmic evolution of various quantities for the model $m(r)=-0.2(1+r)(3.2+0.8r+r^{2})$
with initial conditions $x_{1}=10^{-10}$, $x_{2}=-10^{-7}$, $x_{3}=1.000007\times10^{-7}$
and $x_{4}=0.999$ at the redshift $z=3.5\times10^{6}$. The model
has an approximate matter dominated epoch followed by a non-phantom
accelerated universe with $w_{{\rm eff}}\approx-0.935$. }

\label{spemodel} 
\end{figure}

This model belongs to the Class IV, so the cosmological trajectories
can be acceptable. In Fig.~\ref{spemodel} we find that the matter
epoch is in fact followed by a stable acceleration with $w_{{\rm eff}}\approx-0.935$.
The transition between the various eras is not very sharp compared
to the $\Lambda$CDM model, so it is of interest to investigate in
more detail whether this model can be really compatible with observations.
However this is beyond the scope of this paper.

\section{Conclusions}

The $f(R)$ dark energy models are interesting and quite popular attempts
to explain the late-time acceleration. However it was recently found
that the popular model $f(R)=R+\alpha R^{-n}$ with $n>0$ is unable
to produce a matter era prior to the accelerated epoch \cite{APT}.
In this paper we have attempted to clarify the conditions under which
$f(R)$ dark energy models are cosmologically viable. We first derived
the autonomous equations (\ref{N1})-(\ref{N4}) which are applicable
to general $f(R)$ models. In Sec.~III all fixed points are derived
in such an autonomous system. By considering linear perturbations
about the fixed points, we have studied their stabilities to understand
the cosmological evolution in $f(R)$ dark energy models.

The main result of this paper is that we have identified four classes
of $f(R)$ models, depending on the existence of a standard or {}``wrong''
matter era ($\phi$MDE) and on the final acceleration. In practice,
we have shown that the cosmology of $f(R)$ models can be based on
a study of the $m(r)$ curves in the $(r,m)$ plane and on its intersections
with the critical line $m=-r-1$. This provides an extremely simple
method to investigate the cosmological viability of such models. In
particular, we find that the Class I models correspond to the type
of models in which the final acceleration is preceded by a so-called
$\phi$MDE phase characterized by $a\propto t^{1/2}$ or in which
the matter phase does not exist at all prior to the accelerated epoch.
These models are clearly ruled out, e.g. by the angular diameter distance
of the CMB acoustic peaks, see Ref.~\cite{APT}. This is by far the
largest class and only a few special cases belong to other three.

The general conditions for a successful $f(R)$ model can be summarized
as follows:

\begin{itemize}
\item A $f(R)$ model has a standard matter dominated epoch only if it satisfies
the conditions \begin{equation}
m(r)\approx+0\quad{\rm and}\quad m'(r)>-1\quad{\rm at}\quad r\approx-1\,,\end{equation}
 where the second condition is required to leave the matter era for
the late-time acceleration. 
\item The matter epoch is followed by a de-Sitter acceleration ($w_{{\rm eff}}=-1$)
only if \begin{equation}
0\le m(r)\le1\quad{\rm at}\quad r=-2\quad{\rm or}\quad m(r)=-r-1\to\pm\infty\quad({\rm Class~II})\,.\end{equation}

\item The matter epoch is followed by a non-phantom accelerated attractor
($w_{{\rm eff}}\ge-1$) only if $m=-r-1$ and \begin{equation}
(\sqrt{3}-1)/2<m(r)\le1\quad{\rm and}\quad m'(r)<-1\quad({\rm Class~IV})\,.\end{equation}

\end{itemize}
Moreover, the curve $m(r)$ must connect with continuity the vicinity
of the matter point $P_{5}:(r,m)=(-1,0)$ with one of the accelerated
regions. The Class II and IV models are characterized by $m(r)$ curves
that satisfy these requirements and lead therefore to an acceptable
cosmology.

In the Class III models the curve $m(r)$ intersects the critical
line at $m$ small and negative. In this case the saddle eigenvalue
takes a very large real value and the matter era is practically unstable
and therefore generically very short. Moreover, most trajectories
will be attracted by the strongly phantom attractor with $w_{{\rm eff}}<-7.6$
which is in contrast with observations.

The cases with $m'(r)=-1$ or $m'(r)\to\pm\infty$ at the critical
points are not covered in our linear approach and a higher-order or
numerical analysis is necessary. Also the power-law model $f(R)=R^{-n}$
is a rather special case in a sense that it gives a transition from
the quasi matter era to the strongly phantom epoch with a \textit{constant}
negative $m$. However we showed that this model is not cosmologically
acceptable because of the absence of the prolonged matter epoch. We
have also studied analytically and numerically models like $f(R)=R+\alpha R^{-n},R^{p}(\log\alpha R)^{q},R^{p}\exp(qR),R^{p}\exp(q/R),(R^{a}-\Lambda)^{b}$
and others and have confirmed the conclusions drawn from the $m(r)$
approach. See Table I for the summary of the classification of a sample
of $f(R)$ dark energy models.

As we have seen, the variable $m=Rf_{,RR}/f_{,R}$ plays a central
role to determine the cosmological viability of $f(R)$ dark energy
models. The $\Lambda$CDM model, $f(R)=R-\Lambda$, corresponds to
$m=0$ at all times, which thus satisfies the condition for the existence
of the matter era ($m\approx0$) followed by the de-Sitter point at
$m(r=-2)=0$. The difference from the line $m=0$ characterises the
deviation from the $\Lambda$CDM model. If the devaition from $m=0$
is small, it is expected that such models are cosmologically viable.

We conclude with a comment concerning a possible signature of $f(R)$
cosmology. The standard matter era can be realized with $m\to\pm0$.
As we have seen, in all successful cases we analyzed in this work,
the matter era is realized through damped oscillations with positive
$m$. This raises the obvious question of whether such oscillations
are observable and whether they could be taken as a signature of modified
gravity. This question is left to future work. An additional interesting
direction to investigate is the evolution of cosmological perturbations
in $f(R)$ dark energy models in order to confront with the datasets
of CMB and large scale structure along the lines of Refs.~\cite{Song,Bean,FTBM06}.


\section*{ACKNOWLEDGEMENTS}

We thank Alexei Starobinsky for useful discussions. S.\,T. is supported
by JSPS (Grant No.30318802). L.A. thanks Elena Magliaro for collaboration
in an early stage of this project.


\begin{thebibliography}{10}
\bibitem{review}V.~Sahni and A.~A.~Starobinsky, 
Int.\ J.\ Mod.\ Phys.\ D \textbf{9}, 373 (2000); S.~M.~Carroll,
Living Rev.\ Rel.\  \textbf{4}, 1 (2001); T.~Padmanabhan, 
Phys.\ Rept.\  \textbf{380}, 235 (2003); P.~J.~E.~Peebles and
B.~Ratra, 
Rev.\ Mod.\ Phys.\  \textbf{75}, 559 (2003); V.~Sahni, 
Lect.\ Notes Phys.\  \textbf{653}, 141 (2004) {[}arXiv:astro-ph/0403324].

\bibitem{CST} E.~J.~Copeland, M.~Sami and S.~Tsujikawa, 
Int.\ J.\ Mod.\ Phys.\  D \textbf{15}, 1753 (2006).

\bibitem{SN}S.~Perlmutter \textit{et al.}, 
Astrophys.\ J.\  \textbf{517}, 565 (1999); A.~G.~Riess \textit{et
al.}, 
Astron.\ J.\  \textbf{116}, 1009 (1998); Astron.\ J.\  \textbf{117},
707 (1999); J.~L.~Tonry \textit{et al.}, 
Astrophys.\ J.\  \textbf{594}, 1 (2003); R.~A.~Knop \textit{et
al.}, Astrophys.\ J.\  \textbf{598}, 102 (2003).

\bibitem{CMB}D.~N.~Spergel \textit{et al.}, Astrophys.\ J.\ Suppl.\ 
\textbf{148}, 175 (2003); D.~N.~Spergel \textit{et al.}, 
arXiv:astro-ph/0603449.

\bibitem{LSS}M.~Tegmark \textit{et al.}, 
Phys.\ Rev.\ D \textbf{69}, 103501 (2004); U.~Seljak \textit{et
al.}, Phys.\ Rev.\ D \textbf{71}, 103515 (2005).

\bibitem{BAO}D.~J.~Eisenstein \textit{et al.}, Astrophys.\ J.\ 
\textbf{633}, 560 (2005); C.~Blake, D.~Parkinson, B.~Bassett, K.~Glazebrook,
M.~Kunz and R.~C.~Nichol, 
Mon.\ Not.\ Roy.\ Astron.\ Soc.\  \textbf{365}, 255 (2006).

\bibitem{WL}B.~Jain and A.~Taylor, 
Phys.\ Rev.\ Lett.\  \textbf{91}, 141302 (2003).

\bibitem{Sel} U.~Seljak, A.~Slosar and P.~McDonald, 
JCAP \textbf{0610}, 014 (2006).

\bibitem{Astier}P.~Astier \textit{et al.}, 
Astron.\ Astrophys.\  \textbf{447}, 31 (2006).

\bibitem{CP01} M. Chevallier and D. Polarski, Int. J. Mod. Phys.
D \textbf{10}, 213 (2001).

\bibitem{SS06} V.~Sahni and A.~A.~Starobinsky, arXiv:astro-ph/0610026.

\bibitem{quin}Y.~Fujii, Phys.\ Rev.\ D \textbf{26}, 2580 (1982);
L.~H.~Ford, 
Phys.\ Rev.\ D \textbf{35}, 2339 (1987); C.~Wetterich, Nucl. \ Phys
\ B. \textbf{302}, 668 (1988); B. Ratra and J. Peebles, Phys. \ Rev
\ D \textbf{37}, 321 (1988); Y.~Fujii and T.~Nishioka, 
Phys.\ Rev.\ D \textbf{42}, 361 (1990); E. J. Copeland, A. R. Liddle,
and D. Wands, Ann. N. Y. Acad. Sci. \textbf{688}, 647 (1993); C. Wetterich,
A\&A 301, 321 (1995); P.~G.~Ferreira and M.~Joyce, 
Phys.\ Rev.\ Lett.\  \textbf{79}, 4740 (1997); Phys.\ Rev.\ D
\textbf{58}, 023503 (1998); R.~R.~Caldwell, R.~Dave and P.~J.~Steinhardt,
Phys.\ Rev.\ Lett.\  \textbf{80}, 1582 (1998); I.~Zlatev, L.~M.~Wang
and P.~J.~Steinhardt, 
Phys.\ Rev.\ Lett.\  \textbf{82}, 896 (1999); P.~J.~Steinhardt,
L.~M.~Wang and I.~Zlatev, 
Phys.\ Rev.\ D \textbf{59}, 123504 (1999).

\bibitem{coupled}L.~Amendola, 
Phys.\ Rev.\ D \textbf{62}, 043511 (2000); L.~Amendola and D.~Tocchini-Valentini,
Phys.\ Rev.\ D \textbf{64}, 043509 (2001); L.~Amendola and C.~Quercellini,
Phys.\ Rev.\ D \textbf{68}, 023514 (2003).

\bibitem{AQTW}L.~Amendola, M.~Quartin, S.~Tsujikawa and I.~Waga,
Phys.\ Rev.\ D \textbf{74}, 023525 (2006).

\bibitem{Mel}A.~Melchiorri, L.~Mersini-Houghton, C.~J.~Odman
and M.~Trodden, 
Phys.\ Rev.\ D \textbf{68}, 043509 (2003); U.~Alam, V.~Sahni,
T.~D.~Saini and A.~A.~Starobinsky, 
Mon.\ Not.\ Roy.\ Astron.\ Soc.\  \textbf{354}, 275 (2004); B.~A.~Bassett,
P.~S.~Corasaniti and M.~Kunz, 
Astrophys.\ J.\  \textbf{617}, L1 (2004).

\bibitem{phantom}R.~R.~Caldwell, Phys. \ Lett. \ B \textbf{545},23-29
(2002); R.~R.~Caldwell, M.~Kamionkowski and N.~N.~Weinberg, Phys.\ Rev.\ Lett.\ \textbf{91},
071301 (2003); S.~M.~Carroll, M.~Hoffman and M.~Trodden, Phys.\ Rev.\ D
\textbf{68}, 023509 (2003); P.~Singh, M.~Sami and N.~Dadhich, Phys.\ Rev.\ D
\textbf{68} 023522 (2003).

\bibitem{Cline}J.~M.~Cline, S.~Jeon and G.~D.~Moore, Phys.\ Rev.\ D
\textbf{70}, 043543 (2004); N.~Arkani-Hamed, H.~C.~Cheng, M.~A.~Luty
and S.~Mukohyama, JHEP \textbf{0405}, 074 (2004); F.~Piazza and
S.~Tsujikawa, JCAP \textbf{0407}, 004 (2004).

\bibitem{nonmin}J.~P.~Uzan, Phys.\ Rev.\ D \textbf{59}, 123510
(1999); L.~Amendola, Phys.\ Rev.\ D \textbf{60}, 043501 (1999);
T.~Chiba, Phys.\ Rev.\ D \textbf{60}, 083508 (1999);

\bibitem{stensor} Y. Fujii, Phys. Rev. D\textbf{62}, 044011 (2000);
N.~Bartolo and M.~Pietroni, Phys.\ Rev.\ D \textbf{61} 023518
(2000); F.~Perrotta, C.~Baccigalupi and S.~Matarrese, Phys.\ Rev.\ D
\textbf{61}, 023507 (2000); A. Riazuelo \& J.-P. Uzan, Phys.Rev. D66
(2002) 023525

\bibitem{Esp} G. Esposito-Far\`{e}se and D. Polarski, Phys.\ Rev.\ D
\textbf{63} 063504 (2001).

\bibitem{BEPS00}B. Boisseau, G. Esposito-Far\`{e}se, D. Polarski
and A.~A.~Starobinsky, Phys.\ Rev. Lett. \textbf{85}, 2236 (2000).

\bibitem{GPRS06}R.~Gannouji, D.~Polarski, A.~Ranquet and A.~A.~Starobinsky,
JCAP \textbf{0609}, 016 (2006).

\bibitem{Peri}L.~Perivolaropoulos, JCAP \textbf{0510}, 001 (2005);
S.~Nesseris and L.~Perivolaropoulos, arXiv:astro-ph/0610092.

\bibitem{Capo}S.~Capozziello, V.~F.~Cardone, S.~Carloni and A.~Troisi,
Int.\ J.\ Mod.\ Phys.\ D \textbf{12}, 1969 (2003).

\bibitem{Carroll}S.~M.~Carroll, V.~Duvvuri, M.~Trodden and M.~S.~Turner,
Phys.\ Rev.\ D 70, 043528 (2004).

\bibitem{coa}S.~Capozziello, F.~Occhionero and L.~Amendola, Int.\ J.\ Mod.\ Phys.\ D
\textbf{1} (1993) 615.

\bibitem{Chiba}T.~Chiba, 
Phys.\ Lett.\ B \textbf{575}, 1 (2003).

\bibitem{Dolgov}A.~D.~Dolgov and M.~Kawasaki, Phys.\ Lett.\ B
\textbf{573}, 1 (2003).

\bibitem{Faraoni} V.~Faraoni, 
Phys.\ Rev.\ D \textbf{74}, 023529 (2006).

\bibitem{NO03}S.~Nojiri and S.~D.~Odintsov, Phys.\ Rev.\ D \textbf{68},
123512 (2003).

\bibitem{Brook} A.~W.~Brookfield, C.~van de Bruck and L.~M.~H.~Hall,
Phys.\ Rev.\ D \textbf{74}, 064028 (2006).

\bibitem{nva}I.~Navarro and K.~Van Acoleyen, 
arXiv:gr-qc/0611127.

\bibitem{fRpapers} F.~Perrotta and C.~Baccigalupi, 
 Phys.\ Rev.\  D \textbf{65}, 123505 (2002); S.~Nojiri and S.~D.~Odintsov,
Gen.\ Rel.\ Grav.\  \textbf{36}, 1765 (2004); M.~E.~Soussa and
R.~P.~Woodard, 
Gen.\ Rel.\ Grav.\  \textbf{36}, 855 (2004); G.~Allemandi, A.~Borowiec
and M.~Francaviglia, 
Phys.\ Rev.\ D \textbf{70}, 103503 (2004); D.~A.~Easson, 
Int.\ J.\ Mod.\ Phys.\ A \textbf{19}, 5343 (2004); S.~M.~Carroll,
A.~De Felice, V.~Duvvuri, D.~A.~Easson, M.~Trodden and M.~S.~Turner,
Phys.\ Rev.\ D \textbf{71}, 063513 (2005); S.~Carloni, P.~K.~S.~Dunsby,
S.~Capozziello and A.~Troisi, 
Class.\ Quant.\ Grav.\  \textbf{22}, 4839 (2005); S.~Capozziello,
V.~F.~Cardone and A.~Troisi, 
Phys.\ Rev.\ D \textbf{71}, 043503 (2005); G.~Cognola, E.~Elizalde,
S.~Nojiri, S.~D.~Odintsov and S.~Zerbini, 
JCAP \textbf{0502}, 010 (2005); S.~Nojiri, S.~D.~Odintsov and S.~Tsujikawa,
Phys.\ Rev.\ D \textbf{71}, 063004 (2005); M.~C.~B.~Abdalla,
S.~Nojiri and S.~D.~Odintsov, 
arXiv:hep-th/0601213; R.~P.~Woodard, 
arXiv:astro-ph/0601672; S.~Das, N.~Banerjee and N.~Dadhich, 
Class.\ Quant.\ Grav.\  \textbf{23}, 4159 (2006); S.~Capozziello,
V.~F.~Cardone, E.~Elizalde, S.~Nojiri and S.~D.~Odintsov, 
Phys.\ Rev.\ D \textbf{73}, 043512 (2006); S.~K.~Srivastava, 
arXiv:astro-ph/0602116; T.~P.~Sotiriou, 
Class.\ Quant.\ Grav.\  \textbf{23}, 5117 (2006); arXiv:gr-qc/0611107;
arXiv:gr-qc/0611158; T.~P.~Sotiriou and S.~Liberati, 
arXiv:gr-qc/0604006; A.~De Felice, M.~Hindmarsh and M.~Trodden,
JCAP \textbf{0608}, 005 (2006); S.~Nojiri and S.~D.~Odintsov, Phys.\ Rev.\ D
\textbf{74}, 086005 (2006); A.~de la Cruz-Dombriz and A.~Dobado,
Phys.\ Rev.\  D \textbf{74}, 087501 (2006); S.~Bludman, 
arXiv:astro-ph/0605198; S.~M.~Carroll, I.~Sawicki, A.~Silvestri
and M.~Trodden, 
New J.\ Phys.\  \textbf{8}, 323 (2006); D.~Huterer and E.~V.~Linder,
Phys.\ Rev.\  D \textbf{75}, 023519 (2007); V.~Faraoni, 
Phys.\ Rev.\  D \textbf{74}, 104017 (2006); 
V.~Faraoni and S.~Nadeau, 
Phys.\ Rev.\  D \textbf{75}, 023501 (2007); X.~h.~Jin, D.~j.~Liu
and X.~z.~Li, 
arXiv:astro-ph/0610854; N.~J.~Poplawski, Phys.\ Rev.\ D \textbf{74},
084032 (2006); arXiv:gr-qc/0610133; A.~F.~Zakharov, A.~A.~Nucita,
F.~De Paolis and G.~Ingrosso, 
Phys.\ Rev.\  D \textbf{74}, 107101 (2006); T.~Chiba, T.~L.~Smith
and A.~L.~Erickcek, 
arXiv:astro-ph/0611867; E.~O.~Kahya and V.~K.~Onemli, 
arXiv:gr-qc/0612026; S.~Fay, R.~Tavakol and S.~Tsujikawa, arXiv:astro-ph/0701479;
M.~Fairbairn and S.~Rydbeck, 
arXiv:astro-ph/0701900; G.~Cognola, M.~Gastaldi and S.~Zerbini,
arXiv:gr-qc/0701138; 
B.~Li and J.~D.~Barrow, arXiv:gr-qc/0701111;
T.~Rador, arXiv:hep-th/0701267; 
I.~Sawicki and W.~Hu, 
arXiv:astro-ph/0702278; S.~Fay, S.~Nesseris and L.~Perivolaropoulos,
arXiv:gr-qc/0703006.

\bibitem{APT}L.~Amendola, D.~Polarski and S.~Tsujikawa, 
arXiv:astro-ph/0603703, Physical Review Letters to appear.

\bibitem{key-2}J.~Barrow and A.~Ottewill, J. Phys. A 16, 2757 (1983).

\bibitem{star}A.~A.~Starobinsky, 
Phys.\ Lett.\ B \textbf{91}, 99 (1980).

\bibitem{Jaichan} J.~c.~Hwang, Astrophys.\ J.\  \textbf{375},
443 (1991).

\bibitem{T02}D.~F.~Torres, 
Phys.\ Rev.\ D \textbf{66}, 043522 (2002).

\bibitem{PR05}D.~Polarski and A.~Ranquet, 
Phys.\ Lett.\ B \textbf{627}, 1 (2005).

\bibitem{Capo06} S.~Capozziello, S.~Nojiri, S.~D.~Odintsov and
A.~Troisi, 
Phys.\ Lett.\ B \textbf{639}, 135 (2006).

\bibitem{APT2} L.~Amendola, D.~Polarski and S.~Tsujikawa, arXiv:astro-ph/0605384.

\bibitem{CCT05}S.~Capozziello, V.~F.~Cardone and A.~Troisi, 
Phys.\ Rev.\ D \textbf{71}, 043503 (2005).

\bibitem{clift} J. Barrow and T. Clifton, Phys. Rev D 72, 103005
(2005).

\bibitem{Song}Y.~S.~Song, W.~Hu and I.~Sawicki, 
arXiv:astro-ph/0610532.

\bibitem{Bean}R.~Bean, D.~Bernat, L.~Pogosian, A.~Silvestri and
M.~Trodden, arXiv:astro-ph/0611321.

\bibitem{FTBM06} T.~Faulkner, M.~Tegmark, E.~F.~Bunn and Y.~Mao,
arXiv:astro-ph/0612569.
\end{thebibliography}
\end{document}